\documentclass[sigconf]{acmart}

\usepackage{flushend}
\usepackage{float}
\usepackage{subfigure}

\AtBeginDocument{%
  \providecommand\BibTeX{{%
    \normalfont B\kern-0.5em{\scshape i\kern-0.25em b}\kern-0.8em\TeX}}}
\usepackage{tabularx}
\usepackage{bbding}
\usepackage{graphicx}
\usepackage{caption}
\usepackage{geometry}
\usepackage{amsmath}
\usepackage{makecell}
\usepackage[commandnameprefix=always]{changes}
\usepackage{float}
\usepackage{color}
\usepackage{booktabs}
\usepackage[normalem]{ulem}
\usepackage{tabu}
\usepackage{hyperref}
\usepackage{xcolor}
\usepackage{appendix}

\begin{document}

\copyrightyear{2026}
\acmYear{2026}
\setcopyright{cc}
\setcctype{by}
\acmConference[CHI '26]{Proceedings of the 2026 CHI Conference on Human Factors in Computing Systems}{April 13--17, 2026}{Barcelona, Spain}
\acmBooktitle{Proceedings of the 2026 CHI Conference on Human Factors in Computing Systems (CHI '26), April 13--17, 2026, Barcelona, Spain}
\acmPrice{}
\acmDOI{10.1145/3772318.3790592}
\acmISBN{979-8-4007-2278-3/2026/04}

\author{Gengchen Cao}
\authornote{These authors contributed equally to the work.}
 \email{gencghencao@gmail.com}
\orcid{0009-0004-6789-2098}
\affiliation{Tsinghua University
\institution{}
\city{Beijing}
\country{China}}

\author{Tianke He}
\authornotemark[1]
 \email{tianke.he@outlook.com}
\orcid{0009-0003-7837-1848}
\affiliation{School of Art, Design and Animation
\institution{Sichuan University of Media and communications}
\city{Chengdu}
\country{China}}

\author{Yixuan Liu}
 \email{liuyixuan20240225@163.com}
\orcid{0009-0004-2628-1194}
\affiliation{City University of Hong Kong
\institution{}
\city{Hong Kong, SAR}
\country{China}}

\author{RAY LC}
\authornote{Correspondences can be addressed to ray.lc@cityu.edu.hk.}
\email{ray.lc@cityu.edu.hk}
\orcid{0000-0001-7310-8790}
\affiliation{
\institution{City University of Hong Kong\\Studio for Narrative Spaces}
\city{Hong Kong, SAR}
\country{China}}

\renewcommand{\shortauthors}{Gong, et al.}

\title[Audience in the Loop]{Audience in the Loop: Viewer Feedback-Driven Content Creation in Micro-drama Production on Social Media}

\begin{abstract}
The popularization of social media has led to increasing consumption of narrative content in byte-sized formats. Such micro-dramas contain fast-pace action and emotional cliffs, particularly attractive to emerging Chinese markets in platforms like Douyin and Kuaishou. Content writers for micro-dramas must adapt to fast-pace, audience-directed workflows, but previous research has focused instead on examining writers’ experiences of platform affordances or their perceptions of platform bias, rather than the step-by-step processes through which they actually write and iterative content. In 28 semi-structured interviews with scriptwriters and writers specialized in micro-dramas, we found that the short-turn-around workflow leads to writers taking on multiple roles simultaneously, iteratively adapting to storylines in response to real-time audience feedback in the form of comments, reposts, and memes. We identified unique narrative styles such as AI-generated micro-dramas and audience-responsive micro-dramas. This work reveals audience interaction as a new paradigm for collaborative creative processes on social media.

\end{abstract} 




\maketitle
\section{Introduction}\label{sec:Intro}

Micro-dramas have emerged as a new mode of byte-sized interaction, extending the logic of short-form engagement on platforms such as Instagram and TikTok into more narrative-oriented forms of entertainment. Unlike typical short videos that prioritize visually catchy moments or meme-driven snippets, micro-dramas adopt a serialized structure: they remain byte-sized for asynchronous, scroll-based consumption, yet unfold through sequential episodes that collectively form a coherent narrative arc \cite{chenFactorsDrivingCitizen2021}. In this hybrid format, each installment delivers rapid emotional hooks and plot progression within seconds, while the series as a whole retains the storytelling continuity characteristic of long-form drama \cite{chengCommentShareTikTok2024,chen2023title}. 

The rapid diffusion of micro-dramas is closely tied to the rise of mobile, vertically oriented short-video platforms. Services such as Douyin and TikTok support hundreds of millions of daily viewers who engage in swipe-based, technology-mediated consumption, creating conditions in which serialized short-form stories can circulate at scale. Their recommendation systems, creation tools, and lightweight distribution channels afford writers the ability to publish and update episodes quickly, while enabling audiences to follow evolving narratives through continuous, low-friction viewing \cite{bartolomeLiteratureReviewVideosharing2023,zannettouAnalyzingUserEngagement2024}. These platform-level affordances have transformed micro-dramas from experimental narrative clips into a widely recognizable form of everyday entertainment. China has become a major locus of this transformation; by 2018, short-video platforms had already reached 648 million users, accounting for nearly 80\% of all internet users nationwide \cite{kayeCoevolutionTwoChinese2021,luFifteenSecondsFame2019,sunDouyinMyNourishment2025}.

\begin{figure*}[t]
    \centering
    \includegraphics[width=\textwidth]{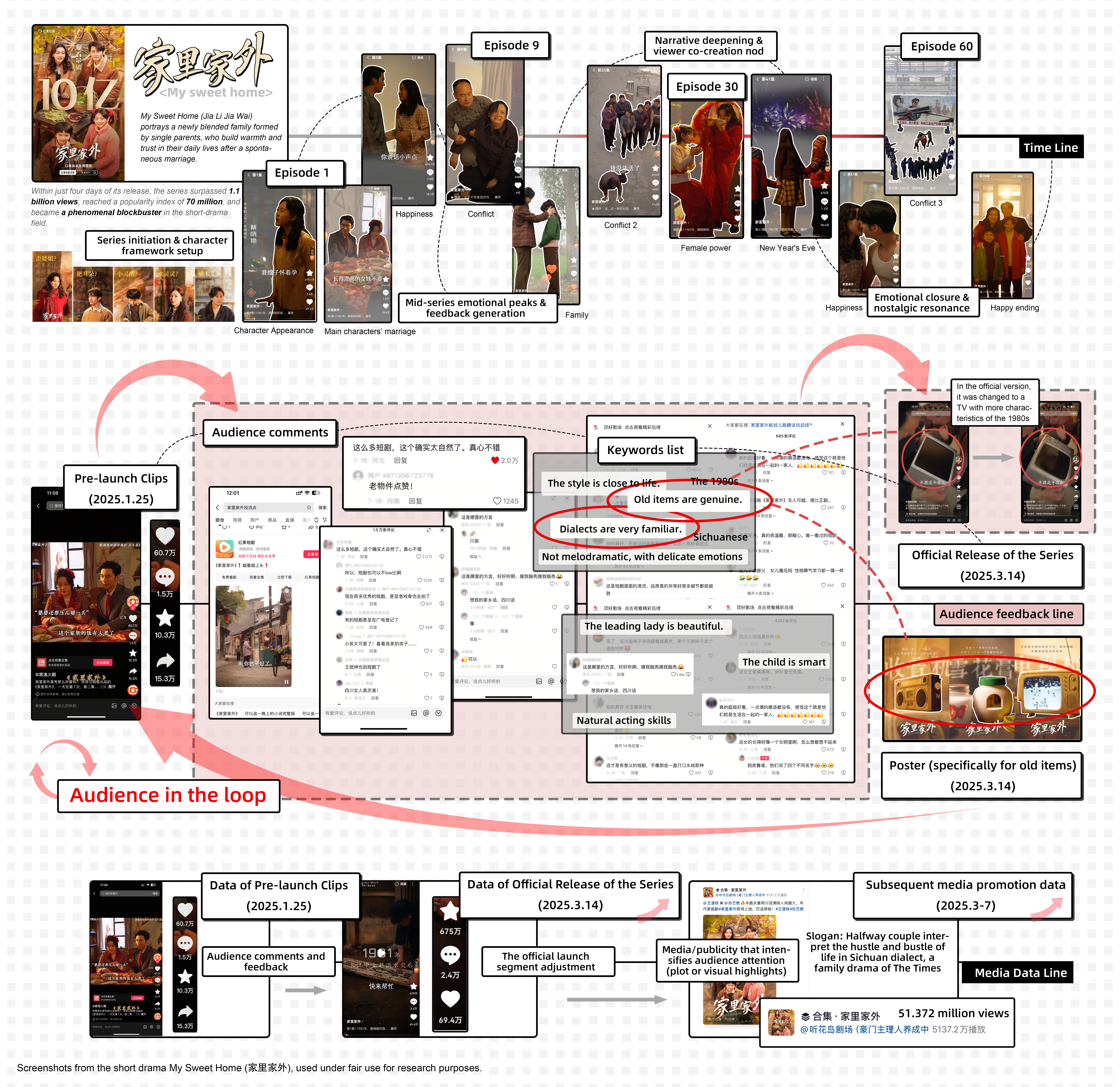}
    \caption{Production workflow of ``My Sweet Home”, showing how audience comments and engagement were integrated into narrative development, promotional design, and media circulation, illustrating an ``audience-in-the-loop” model of micro-drama creation.}
    \label{fig:placeholder}
\end{figure*}

While there are tens of thousands of serialized episodes of micro-drama production released annually on platforms such as Douyin and TikTok, the interactive mechanisms that shape their creation remain underexamined.The current ecology of micro-dramas is fundamentally shaped by the vertical short-video format. Building on this format, prior research has primarily examined user motivations for short-video engagement \cite{kayeCoevolutionTwoChinese2021}, the infrastructuralization of platforms \cite{helmondPlatformizationWebMaking2015,plantinInfrastructureStudiesMeet2018}, and the broader frameworks of new media, short-video platforms, and cultural production \cite{maityUnderstandingPopularitySocial2017,quMicrobloggingMajorDisaster2011}. Yet, little is known about the motivations and workflows of content writers. As an emerging narrative form, micro-dramas are reshaping both aesthetic preferences and interactive practices in the era of social media. Thus, developing a systematic understanding of their production mechanisms is crucial not only for improving platform design but also for advancing digital cultural production and broader cultural development.
We frame this gap not as an industrial analysis nor as a study of writing as an isolated cognitive task, but as an opportunity to understand how writer–audience–platform interactions jointly shape a new mode of serialized, feedback-responsive authorship.

We conducted semi-structured interviews with 28 participants, including micro-drama scriptwriters, content creators, editors working within multi-channel networks (MCNs). MCNs are organizations that partner with creators to provide services such as content optimization, audience analytics, distribution support, and cross-platform coordination, while aggregating creator and writer channels into larger networked entities \cite{gardner2016s}. They function as intermediaries that mediate platform interactions and help creators and writer navigate algorithmic environments \cite{dingUploadersWeHave2022,gardner2016s}. Our sample also included one micro-drama actor. These participants possessed both practical experience and structured insights into feedback-driven production logics. Drawing on empirical accounts, we examined how creators described leveraging signals such as comments, bullet screens, repost metrics, popularity indices, and user-generated tags to dynamically adjust narrative pacing, character relationships, and plot trajectories, resulting in a reflective and feedback-driven mode of authorship.

In addition to sharing their creative routines, participants frequently referenced their own representative works or recent productions as examples. Through these accounts, we gained insight into how micro-dramas were iteratively developed and modified, not through preplanned structures, but through writers’ ongoing responses to platform-driven feedback cycles. Rather than treating micro-dramas as a media industry or screenwriting as an individual cognitive activity, this study focuses on how writers navigate platform-mediated feedback and collaborate with audience responses in everyday creative practice, resulting in a mode of authorship that is shaped through continuous, situated interaction.

This study provides novel insights into how micro-dramas redefine narrative production by embedding implicit, real-time audience feedback into the creative process. We focus on two central questions:

\textbf{RQ1}: \textit{What emerging roles and production workflows have creators adopted in production of micro-dramas?}

\textbf{RQ2}: \textit{How do writers work with audiences to iteratively shape the plots and characters for micro-dramas?}

These questions offer a unique entry point for examining creative practices in the platformized ecosystem of social media.

Our findings suggest that micro-dramas foster a distinctive co-creation mechanism: writers continuously adapt stories and characters in compressed cycles, guided by implicit audience signals such as trending keywords, meme reinterpretations, and repost behaviors. This process not only restructures narrative production but also transforms creative identities, as individuals fluidly shift between roles of writer, actor, director, and editor. Importantly, we show that audience empathy and engagement are more accurately reflected through reposts and tagging practices than through conventional metrics like ``likes''.

Moreover, we conceptualize micro-dramas as data-responsive, market-oriented cultural products, dependent on rapid audience validation, strategic topic selection, and iterative updates. Within this ecology, platform moderators play a pivotal yet understudied role as the ``first audience”: their subjective evaluations and governance protocols both amplify certain creative directions (e.g., emotional hooks, trending topics) and constrain narrative boundaries (e.g., politically sensitive or value-discordant content). This highlights micro-drama production as a form of ``constrained dynamic iteration,” where stories evolve through writer–audience interaction but remain embedded within the limits of platform governance. 

This study makes three contributions.
\begin{itemize}
     \item 
     First, we extend research on technology-mediated creative work by positioning micro-dramas within platform-based narrative systems \cite{riedl2010narrative,ryanNarrativeVirtualReality2015}. We show how implicit audience signals—such as comments, memes, and repost patterns—reshape storytelling not through predefined branches but through continuous, data-responsive feedback loops embedded in sociotechnical infrastructures \cite{helmondPlatformizationWebMaking2015,plantinInfrastructureStudiesMeet2018}. This reframes authorship from a choice-driven model toward an ongoing process of negotiation among writers, audiences, and platform systems.

    \item 
    Second, we extend research on digital writing by showing that micro-drama writers operate as multi-role practitioners whose scripting, performing, directing, and editing unfold within compressed, technology-mediated production cycles \cite{burgess2018youtube,wohnAudienceManagementPractices2020}. Their narrative work is shaped by continuous audience signals and platform governance, producing a form of constrained dynamic iteration in which writing shifts from an author-centered activity to a distributed, co-produced process involving multiple stakeholders.

    \item 
    Third, we outline design opportunities for platforms system. Our results point to the need for systems that help writers interpret audience signals transparently, maintain narrative coherence during rapid iteration, and preserve creative autonomy within writer–audience–platform loops \cite{gomesConvergenceCultureWhere2008,plantinDigitalMediaInfrastructures2019}. Treating micro-dramas as data-responsive cultural artifacts, we identify ways platform design can balance audience influence with narrative diversity, clarify governance constraints, and mitigate creative risks produced by opaque algorithmic mechanisms.
\end{itemize}

\section{Related Work}\label{sec:Related Work}

\subsection{Emergence of Micro-Drama as a Narrative Form}
Micro-dramas are a rapidly expanding form of serialized short-form storytelling characterized by ultra-brief episodes—typically one to three minutes—designed for mobile, vertically oriented viewing. Compared with general online content and short-form videos that emphasize immediate visual impact or standalone moments, micro-dramas unfold through sequential episodes that together build a coherent narrative arc. This hybrid format combines the brevity and immediacy of short-form video with the continuity of serialized drama, making it particularly suited to platforms built around rapid, swipe-based consumption \cite{ernst2022micro,luoDigitalStorytellingNew2024}. To clarify how micro-dramas differ from ordinary short videos and other forms of online content, Figure 2 provides a conceptual comparison across media form, narrative continuity, and interactional features.

\begin{figure*}[t]
    \centering
    \includegraphics[width=0.8\textwidth]{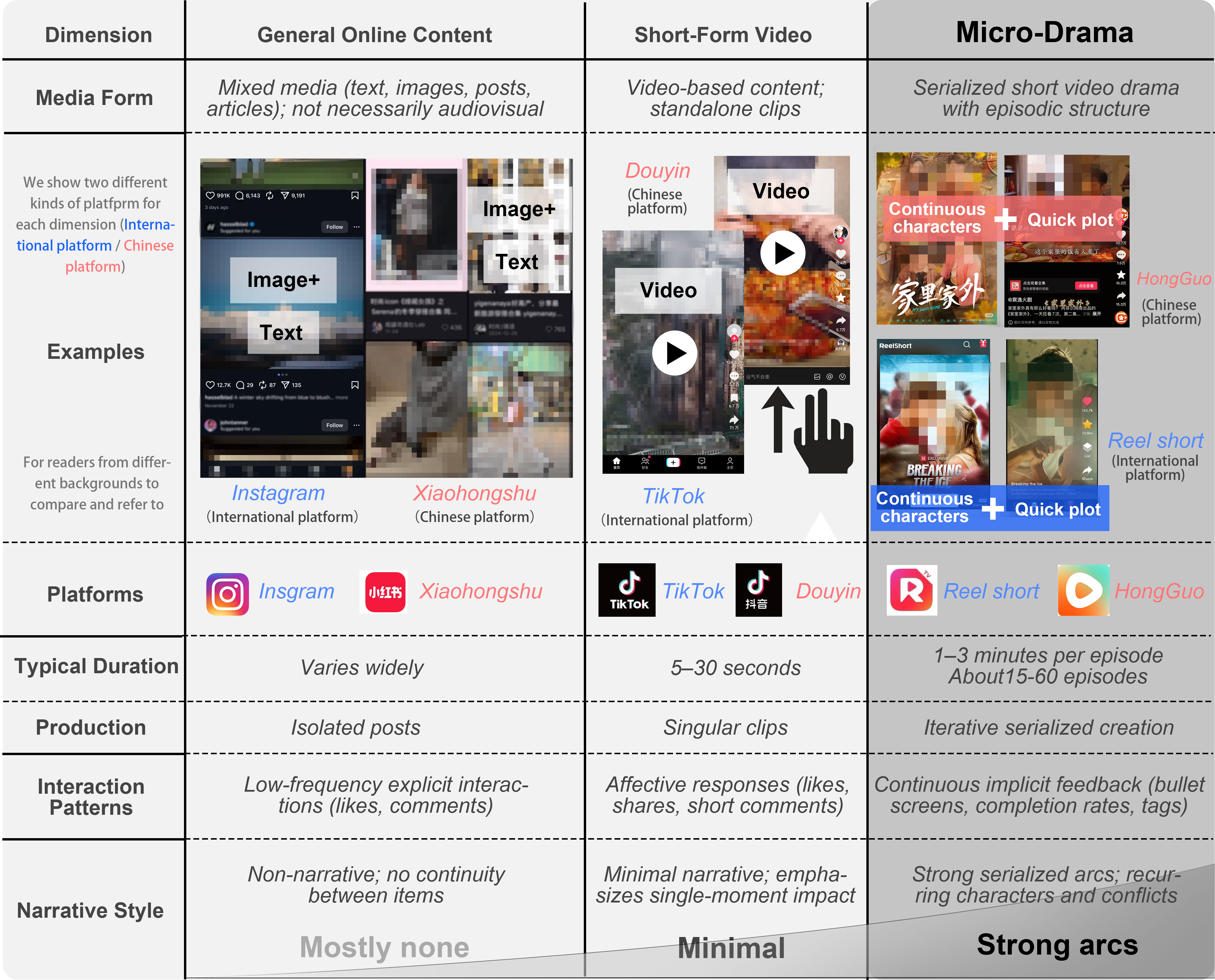}
    \caption{Distinguishing micro-dramas from short-form videos and general online content in platformized media ecosystems.}
    \label{fig:placeholder}
\end{figure*}

The emergence and popularization of micro-dramas are deeply shaped by platform-level affordances. They first gained visibility on global short-video platforms such as TikTok and Instagram \cite{noveriaAnalyzingLanguageFunctions2023}, which support swipe-based consumption of vertically formatted, algorithmically recommended content. In China, platforms such as Douyin \cite{chenStudyCharacteristicsDouyin2019}, a domestic counterpart that offers similar recommendation mechanisms together with extensive writer tools and in-app production workflows, expanded this narrative experimentation. Additional infrastructures such as Hongguo, a dedicated micro-drama application that distributes serialized short dramas through advertisement-supported or pay-per-episode models, further reduced the friction for continuous viewing. Together, these systems provide the technical foundations, distribution mechanisms, and interaction patterns that enable micro-dramas to develop as a distinct narrative form within today’s mobile media environment \cite{ernst2022micro}.

Micro-dramas not only reshape viewing habits but also reorganize the interactional dynamics among audiences, writers, and platform infrastructures. Prior studies on short-form media show that surprise, emotional resonance, and humor are central motivators for engagement \cite{choiProxonaSupportingCreators2025,jinImpactFunExperience2024,wangHumorCameraView2020}. In micro-dramas, these experiential triggers work in tandem with continuous audience traces such as comments, bullet screens, reposts, and tagging behaviors \cite{heShortVideoSegmentlevel2025}. These traces function as implicit signals that circulate within \linebreak technology-mediated feedback loops, allowing writers to sense shifting audience expectations and make narrative adjustments during ongoing production rather than after publication.

\begin{figure*}[t]
    \centering
    \includegraphics[width=\textwidth]{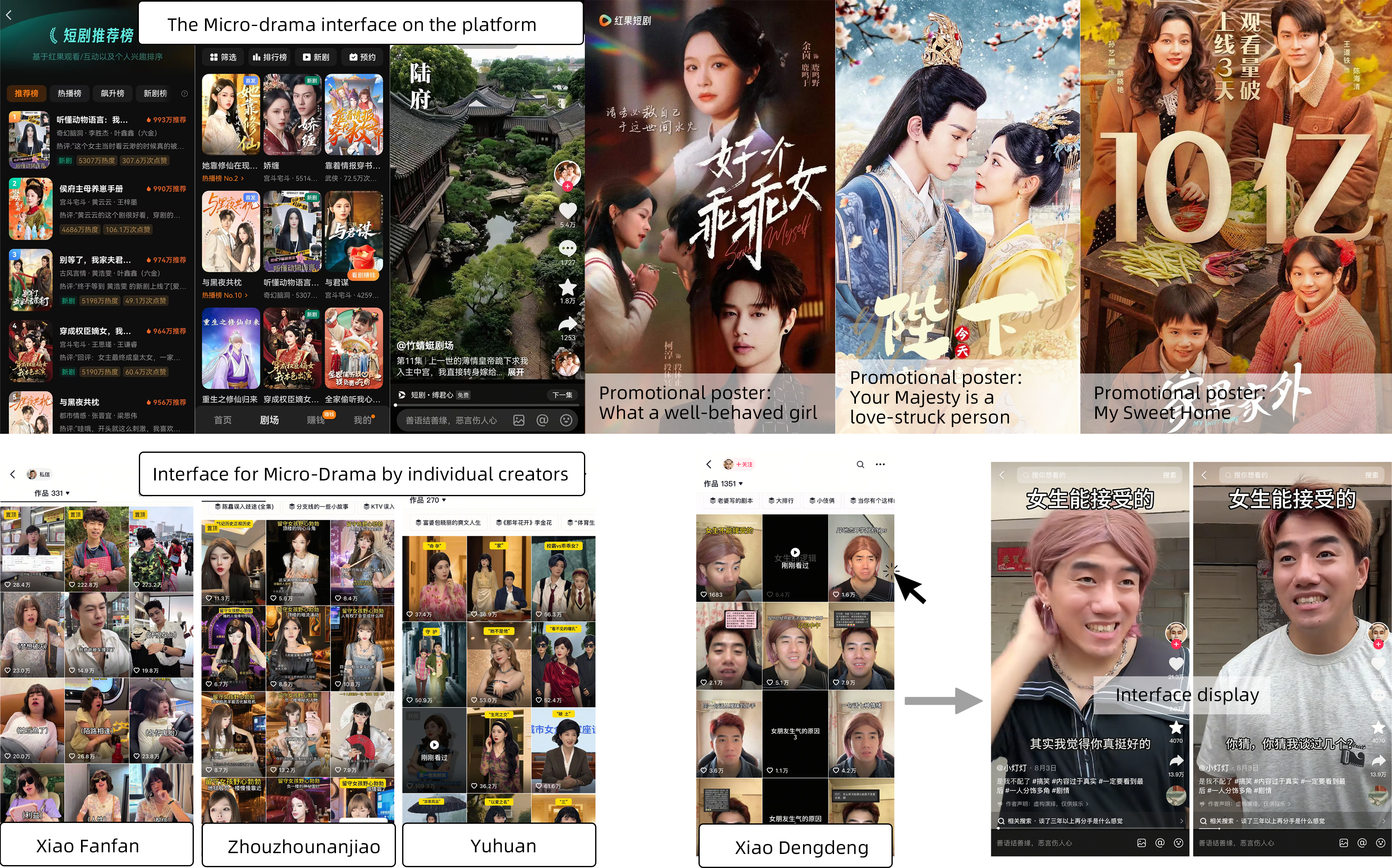}
   \caption{Example of micro-drama interfaces: top row shows platform-level distribution and promotion; bottom row shows individual creators’ production interfaces.}
    \label{fig:placeholder}
\end{figure*}

This environment is also shaped by platform governance. Recommendation mechanisms and content moderation practices serve as early gatekeeping layers that influence visibility and circulation \cite{maHowAdvertiserfriendlyMy2021,heShortVideoSegmentlevel2025}. Research on writer–platform relations shows that such algorithmic systems can reshape creative risk-taking and decision-making by imposing uncertain boundaries that writers must navigate while responding to real-time audience reactions \cite{flores-saviagaAudienceStreamerParticipation2019}. These platform-level structures introduce sociotechnical constraints that differentiate micro-drama production from more traditional narrative media \cite{hodlContentCreatorsPlatform2023}.

Although micro-dramas have expanded rapidly across short-video ecosystems, they remain insufficiently examined within studies of narrative practice, creative labor, and technology-mediated authorship. Understanding how their storytelling processes emerge through iterative adjustments, shifting audience signals, and platform infrastructures requires a closer look at the creative workflows and interactional mechanisms that define this format \cite{chen2025research,bartolomeLiteratureReviewVideosharing2023}. 

\subsection{Interactional and Platform-Shaped Practices in Micro-Drama Writing}
Screenwriting has historically been as conceptualized both an art and a science, with established narrative frameworks such as save the cat beat sheet, the three-act structure \cite{fieldScreenplayFoundationsScreenwriting2005}, and the hero’s journey \cite{campbellHeroThousandFaces2004}providing quantifiable guidance for theater, television, and film. 
Compared with this kind of classic structure and long-form narratives that emphasize complete story arcs, micro-dramas are shaped by platform logics, algorithmic recommendation, and constant audience interaction \cite{bishopManagingVisibilityYouTube2019,cullenNotJustYou2025}. These conditions produce storytelling patterns defined by fast pacing, dense information delivery, and frequent plot reversals \cite{zhang2024ritual}. In this narrative environment, content creator refers to practitioners who often take on multiple responsibilities, including story development, acting, filming, editing, and operational work \cite{simpsonRethinkingCreativeLabor2023,lam2019multi}. This differs from conventional film and television workflows where  writers, directors, and actors are distinct roles. In this study, micro-drama writer refers specifically to the individual who leads narrative construction, pacing, and character development, although their work unfolds within collaborative processes structured by team coordination and platform dynamics.

Within this multi-role creative arrangement, writing is not limited to linguistic expression \cite{chen2025research}.  It functions as a distributed, practice-based, and interaction-oriented narrative activity.  Writers must manage plot progression, visual rhythm, emotional tone, interpretations of audience feedback, and the monitoring of platform signals, forming a continuous cycle of generating, releasing, receiving responses, and adjusting.

However, the highly compressed narrative structure of micro-dramas introduces specific risks. To ensure that audiences can recognize characters within seconds, writer often rely on highly prototypical character templates.
This approach improves narrative efficiency, yet literary studies have long emphasized that the use of flat or stereotypical characters carries structural consequences for how stories generate meaning \cite{forster1927aspects,chen2025research}. Flat characters are constrained by a narrow set of recognizable traits, which reduces their capacity to grow, change, or act in ways that challenge audience expectations \cite{forster1927aspects}. When these restricted patterns dominate a narrative environment, the storytelling process begins to rely on simplified social categories. As a result, representational depth gives way to formulaic depictions that can inadvertently stabilize assumptions about identity, morality, or relational dynamics.

Research on digital storytelling environments expands this critique by showing how stereotype-driven character design becomes amplified under conditions of rapid content circulation. Studies of platform-mediated creativity reveal that algorithms tend to favor content that is immediately legible and emotionally predictable, since such content performs well in early engagement metrics \cite{bishopManagingVisibilityYouTube2019,mcdonald2021powerful}. Under these conditions, flat characters are not simply a creative shortcut. They become part of a broader socio-technical mechanism that rewards formulas and discourages experimentation \cite{forster1927aspects,simpsonRethinkingCreativeLabor2023}. 

The limitations introduced by flat characters also affect audience interpretation. When viewers repeatedly encounter characters who embody narrow traits, they may internalize these portrayals as natural or self-evident. This process can normalize existing social stereotypes and suppress more nuanced or critical forms of representation. Studies on collaborative meaning-making demonstrate that audiences often co-construct character meaning through comments, memes, and shared references \cite{simpsonRethinkingCreativeLabor2023}. When these interpretations center on stereotypical traits, they can further entrench flattened portrayals, making it more difficult for writers to introduce narrative variation without disrupting audience expectations.

Taken together, these developments show that writing in micro-drama narrative content is not solely the outcome of individual creativity but a negotiated practice shaped by platform infrastructures and the evolving interpretations of audiences. The pressures created by compressed formats, algorithmic incentives, and rapid cycles of viewer response encourage storytellers to prioritize legibility and emotional immediacy. At the same time, the circulation of comments, memes, and collective readings establishes a shared interpretive space in which character meaning is continually reconstructed. This environment turns writing into an adaptive process in which writers balance narrative efficiency with representational responsibility while responding to signals that emerge from both audience behavior and platform governance \cite{wu2022participatory}. Understanding these intertwined dynamics clarifies why micro-drama writing tends to converge on recognizable templates, because it reveals the tension among compressed plot design, platform-oriented logic, and the shaping of character representation \cite{chen2025research,liNarrativeAestheticFeatures2024,wu2022participatory}.

\subsection{Audience Participation in Micro-Drama Narratives}
Creative writing, as a text-based form of creative labor, foregrounds the artistic qualities of language, narrative imagination, and stylistic diversity \cite{tangUnderstandingScreenwritersPractices2025}. Within digital platform environments, audience participation has emerged as a critical force reshaping creative workflows, as reflected in research on participatory media \cite{jenkinsTextualPoachersTelevision}, writer–audience communication \cite{naabStudiesUsergeneratedContent2017}, human-AI co-writing \cite{chen_once_2025,yang_ai_2022}, and multi-platform collaboration \cite{maMultiplatformContentCreation2023}, which together provide a foundation for understanding how interaction shapes contemporary creative writing practices.

Research on participatory media and writer–audience communication indicates that audience interactions such as likes, shares, and comments not only shape dissemination outcomes but also drive writers to adapt narrative pacing and visual presentation \cite{dongShortVideoMarketing2024}. These interaction mechanisms simultaneously enhance users’ emotional engagement, resonance, trust, and propensity for further sharing \cite{leeWeBuildAIParticipatoryFramework2019,harringtonDeconstructingCommunitybasedCollaborative2019}.Early work on digital narratives emphasized branching structures and explicit user choices, where audiences could influence the storyline by selecting predefined options \cite{ryanNarrativeVirtualReality2015,riedl2010narrative}. However, this model struggles to capture the latent emotions and preferences expressed through comments, keywords, and emoji reactions on social platforms. Although prior research has examined the influence of such latent signals on recommendation algorithms and content moderation, their systematic integration into narrative generation and editing processes remains underexplored. In contrast, recent studies on live streaming highlight the emergent and productive characteristics of audience feedback. Flores Saviaga \cite{flores-saviagaAudienceStreamerParticipation2019} and Hamilton \cite{hamiltonStreamingTwitchFostering2014} show that audiences shape content direction through real-time comments, tipping, or emoji reactions, redefining performer–audience relationships and framing streaming communities as sites where collective audience behavior modulates performer actions and content pacing. Li et al \cite{liChannelingEnduserCreativity2022} . further demonstrate in game streaming that real-time comment mechanisms enable developers to absorb player feedback instantly, transforming audiences from passive viewers into active collaborators. By comparison, micro-dramas exhibit distinct interaction dynamics: micro-drama writers typically cannot access feedback synchronously and instead rely on discrete cues such as high-frequency comments, trending memes, emotional fluctuations, and branching discussions as latent narrative signals. This non-real-time, latent, and distributed feedback structure renders micro-drama production more dependent on interpreting emergent information and making cross-contextual inferences. Moreover, advances in artificial intelligence have significantly influenced user co-creation, giving rise to two streams of research. The first examines writers’ cognition, identity, and copyright concerns in AI-collaborative contexts \cite{bangerlCreAItiveCollaborationUsers2025,heWhichContributionsDeserve2025,limaPublicOpinionsCopyright2025,han_when_2024, zeng_ronaldos_2025,wanPolymindParallelVisual2025}. The second focuses on the development of collaborative writing and feedback tools, including chart-based text editors \cite{massonTextoshopInteractionsInspired2025,lisnicPlumeScaffoldingText2025}, AI systems supporting feedback and reflection  \cite{brannonAudienceViewAIassistedInterpretation2024}, and large-scale multiuser online co-creation platforms  \cite{leongParatrouperExploratoryCreation2025,mongeroffarelloInvestigatingHowComputer2025,zhangFrictionDecipheringWriting2025}. While these studies enhance the contextuality of creative work and tool dependence, they predominantly emphasize the writer–tool interaction. Even when audience feedback is considered, it is often presented in aggregated or summarized form, leaving underexplored how such feedback is interpreted, prioritized, and translated into concrete creative decisions in everyday practice.

Research on multi-platform practices shows that users increasingly navigate multiple online environments simultaneously \cite{maMultiplatformContentCreation2023,tandocPlatformswingingPolysocialmediaContext2019,zhaoSocialMediaEcology2016} , discussing micro-dramas across platforms such as TikTok, Bilibili, and Xiaohongshu. Rahwan’s society-in-the-loop framework argues that sociotechnical systems should incorporate public values into their operational cycles to support socially responsive decision-making \cite{rahwanSocietyintheloopProgrammingAlgorithmic2018} . In content production, heterogeneous platform algorithms and operational goals jointly shape writers’ decision environments, requiring them to integrate fragmented signals from multiple platforms when refining plot structures and character arcs \cite{walshAICodesignDeveloping2019,banks2009co}. Micro-drama writers similarly rely on cross-platform emotional cues and audience reactions, treating broader social feedback as a significant driver of narrative direction. Despite this reliance, writers still process multi-platform comments manually, resulting in severe fragmentation, limited structure, and no technical support for synthesizing heterogeneous signals. Prior systems work, such as Hong et al.’s  \cite{hongFosteringCollectiveDiscourse2025a} Distributed Role-Based Online News Commenting System, has demonstrated how fragmented comments can be clustered, summarized, and structured to support collective sensemaking. However, the emotional, fast-paced, and entertainment-oriented nature of micro-drama engagement makes such systems incompatible with the semantic assumptions, user motivations, and temporal rhythms of this domain. Platform-specific differences further complicate feedback integration. As shown by Ma et al. \cite{maMultiplatformContentCreation2023} and DeVito et al. \cite{devitoPlatformsPeoplePerception2017} , variations in platform structures, algorithms, and cultures produce distinct patterns of audience expression. Manual browsing cannot identify the underlying value differences across platforms or align feedback across contexts, making it difficult for writers to extract decision-relevant signals from large volumes of heterogeneous comments. Existing research has yet to examine how writers integrate audience participation across platforms or how platform-diverse feedback can be transformed into actionable narrative adjustments. These gaps indicate a broader theoretical and methodological need to understand how cross-platform audience participation shapes micro-drama content production.

In summary, although existing literature offers valuable insights into user participation and platform mechanisms, it has yet to systematically address how writers identify, align, and translate high-value latent audience signals into actionable narrative adjustments within non-real-time, cross-platform, and high-volume comment environments; how audiences, platform algorithms, and writers collectively form a dynamic narrative regulation loop; and how these processes shape narrative design, pacing, and content iteration in micro-dramas. Through in-depth interviews with micro-drama writers, this study aims to fill these theoretical and methodological gaps and reveal how audience feedback, mediated by platform mechanisms and writer agency, reshapes contemporary short-form narrative production practices.

\subsection{The Multiple Constraints on Creative Labor in Micro-Drama Production}
Platforms such as YouTube, TikTok, and Hongguo have fueled the rise of the “writer economy” \cite{
ryanNarrativeVirtualReality2015,sanyouraQuantifyingCreatorEconomy2022,maMultiplatformContentCreation2023}, enabling writers to reach large-scale audiences and convert content production into economic value. Recent scholarship has introduced socio-economic perspectives to examine issues of creative labor \cite{banks2009co,duffyNestedPrecaritiesCreative,hesmondhalghVeryComplicatedVersion2010}and relational labor \cite{bonifacioFansRelationalLabor2023,duffyNestedPrecaritiesCreative}. Within this context, platformized creative labor is conceptualized as a multilayered process shaped by algorithmic governance, audience feedback, and technological tools. It encompasses not only content production but also the maintenance of audience relationships, the management of self-presentation, and negotiation with platform rules and infrastructural constraints \cite{simpsonRethinkingCreativeLabor2023}.

In this study, we adopt Simpson \cite{simpsonRethinkingCreativeLabor2023}and colleagues’ conceptualization of creative labor, which emphasizes that writers must continuously engage in relational and visibility labor to maintain exposure and sustain the stability of their creative ecosystems \cite{bishopManagingVisibilityYouTube2019,cullenNotJustYou2025} .First, algorithmic preferences and content norms impose structural constraints on creative expression. Ma et al. \cite{maHowAdvertiserfriendlyMy2021} show that opaque algorithmic moderation systems force writers to invest substantial risk-management labor, including avoiding penalties, handling appeals, and responding to unpredictable takedowns. Such labor is invisible, unpredictable, and typically uncompensated. In micro-drama production, where episodes are short and update cycles are rapid, writers must not only generate narratives that align with algorithmic recommendation logic but also comply with platform requirements regarding plot conflicts, emotional hooks, and fixed episode lengths. These conditions intensify time pressure, constrain creative exploration, and heighten psychological burden.

Second, the pursuit of visibility often pushes writers toward narrow themes, formulaic content structures, and stable persona management, which restricts expressive range and accelerates the homogenization and commodification of creative work, contributing to writers' sense of alienation. To match platform algorithms and fast-paced audience feedback, micro-drama writers frequently converge on a limited set of high-traffic genres—such as “rebirth revenge,” “CEO romance,” or “family conflict”—and rely on 1–3-minute episodes built around rapid emotional hooks. Characters are often reduced to archetypes like the “evil supporting female,” “domineering CEO,” or “fragile heroine,” with minimal psychological depth or coherent motivations. Repeated production within a data-driven framework gradually erodes writers’ narrative agency, further reinforcing burnout and creative detachment. As Kim et al. \cite{huExploringDanmakuContent2024} emphasize, even AI tools such as ASVG can increase production efficiency yet struggle to preserve writers’ unique styles or support empathetic communication, potentially reinforcing existing algorithmic narrative path dependencies, increasing uncertainty, and limiting expressive autonomy.

Finally, at the writer–audience level, platforms reinforce structural asymmetries: writers shoulder intensive labor while audience participation carries virtually no cost. Metrics such as watch time, comments, and clicks are transformed by algorithms into implicit signals that steer production \cite{simpsonYouForyouEveryday2021}. Micro-drama writers must simultaneously perform emotional labor and adapt to algorithmically generated “audience profiles,” which heightens production pressure, emotional exhaustion, and a sense of distance from their audience \cite{bonifacioFansRelationalLabor2023,cunningham2019social}. These tensions have prompted researchers to explore technological interventions. For example, Proxona generates data-driven audience personas to reduce writers’ cognitive gaps with their audiences \cite{choiProxonaSupportingCreators2025} . However, in fast-paced, multimodal, and narrative-dense micro-drama ecosystems, such tools still face challenges including delayed feedback, text-centric biases, and limited capacity to meaningfully translate audience labor into actionable creative support, resulting in labor misalignment.

Overall, the rise of the writer economy has positioned digital writers as central cultural and economic actors whose production, distribution, and monetization systems are deeply dependent on audience participation \cite{bhargavaCreatorEconomyManaging2022,harriganIdentifyingInfluencersSocial2021,johnsonIntroducingContentpreneurMaking2022,tafesse2023content}. While prior work has extensively examined creative labor in livestreaming and short-video environments, micro-dramas—characterized by strong narrative orientation and highly feedback-driven production cycles—remain underexplored. Situating micro-drama within the frameworks of creative labor and the writer economy enables a closer examination of the power relations and labor tensions embedded in narrative production, and underscores the need for platform design that safeguards both creative value and labor dignity to support the sustainable development of emerging narrative ecologies.

\begin{table*}[t]
\vspace{-0.2cm}
\caption{Demographic Information of Participants}
    \centering
    \resizebox{\textwidth}{!}{%
    \begin{tabular}{|c|c|c|c|c|c|}
    \hline
        \textbf{ID} & \textbf{Gender} & \textbf{position} & \textbf{platform} & \textbf{seniority}  \\
        \hline
       P1&  Female&  Micro-drama Screenwriter&  Hongguo, Douyin&  
5 years\\
 P2&  Female& Micro-drama Editor,Micro Drama Content Creator&  Hongguo, Douyin&  
3 years\\
 P3&  Male& Micro-drama screenwriter,director&  Douyin&  
6 years\\
 P4&  Male& Micro-drama Screenwriter&  TikTok,YouTube&  
2 years\\
P5&  Female& Micro-drama Screenwriter&  Douyin&  
1 year\\
P6&  Female& Micro-drama Screenwriter&  Overseas Micro Drama Platform&  
1 year\\
P7&  Female& Micro-drama Screenwriter&  Overseas Micro Drama Platform&  
1 year\\
P8&  Female&Micro-drama Director&  Douyin, Video Account&  1 year\\
p9&  Male&Micro-drama Director&  Video Account, Shipinghao&  1 year\\
p10&  Male&Micro-drama Director&  Hongguo, Douyin&  6 years\\
p11& Female&Micro-drama Director& Overseas Micro Drama Platform,TikTok,YouTube& 6 months\\
p12& Female&Micro-drama Screenwriter& Tencent& 3 years\\
p13& Female&Micro-drama Screenwriter& Douyin& 9 months\\
p14& Female&Micro-drama Producer& Douyin& 2 years\\
p15& Male&Micro-drama Actor& Douyin& 2 years\\
p16&Female&Part-time Micro-drama Screenwriter,TV Drama Screenwriter, Online Novelist& Tencent,iQiyi& 6 years\\
p17&Female&Part-time Micro-drama Screenwriter,TV Drama Screenwriter& Cooperative Film,iQiyi,TV Platform& 3 years\\
p18&Male&Micro-drama Script Supervisor,Theater Drama Screenwriter, Game Planner&Cooperative Film& 3 years\\
p19&Female&Micro-drama Script Producer,Theater Drama Screenwriter&Douyin&3 years\\
p20&Male&Micro-drama Screenwriter&Douyin&3 years\\
p21&Male&Micro-drama Screenwriter&Douyin&2 years\\  
p22&Male&Micro-drama Director&Hongguo&3 years\\ 
p23&Female&Micro-drama actor&Hongguo&1 year\\
p24&Female&animated Micro-drama director&Netease&1 year\\
p25&Female&Micro-drama Producer &Dramabox&1 year\\
p26&Male&Micro-drama Screenwriter&Dramabox&1 year\\
p27&Female&Micro-drama Screenwriter&Tencent&3years\\
p28&Male&Micro-drama MCN manager&Tencent&5years\\
        \hline
    \end{tabular}%
    }
    \label{tab:my_label}
\end{table*}

\section{Methods}\label{sec:Methods}

This study aims to gain a deeper understanding of how micro-drama writers incorporate audience feedback into their scriptwriting process. We explored RQS through semi-structured interviews and qualitative analysis of interview transcripts.

\subsection{Data collection: Interviews}
\textbf{3.1.1 Participant recruitment}
We recruited participants through acquaintance referrals and snowball sampling. The first batch of participants were colleagues of research team members. Subsequently, we contacted more participants through referrals from interviewees. All participants needed to meet the following three criteria: 1. Participants must have at least six months of micro-drama creation experience and be part of a writing team rather than individual writer. 2. Participants' works must have been released on short video platforms or short drama platforms, or have received awards related to micro-drama. 3. Traditional screenwriters must have at least one officially broadcast work to allow for comparative analysis with micro-drama.

\textbf{3.1.2 Semi-Structured Interviews}
Interviews were conducted through Tencent Meeting or WeChat voice chat. During each interview, one researcher primarily led the questioning and follow-up probes, while another researcher took detailed notes and monitored emerging themes. All interviews were audio-recorded in Mandarin. Each research question (RQ) was examined using specifically designed interview questions. These questions were deliberately open-ended to elicit detailed accounts of the mechanisms and practices participants engaged in during the creative process(see Appendix A).

Each interview followed a consistent process: participants first provided informed consent, completed a short background survey, and introduced their professional background; this was followed by semi-structured questions about their writing practices, opportunities to share representative works or platform profiles, and a brief debriefing session in which participants could raise additional questions or reflections. Interviewees frequently shared experiences beyond the predefined questions. To capture such emergent practices, we kept the interviews semi-structured, allowing room for follow-up questions that often revealed insights into their creative workflows.

\subsection{Participants}
This study recruited 28 participants, including 24 micro-drama practitioners (screenwriters, directors, producers, actors, etc.) and 4 traditional screenwriters. The gender distribution was 17 female and 11 male, with work experience ranging from 6 months to 9 years, covering major domestic and foreign platforms such as Douyin, TikTok, and Tencent.Due to the highly fluid and multifunctional nature of roles in micro-drama creation, micro-drama practitioners typically work in small teams to complete the entire process from planning, scriptwriting, and performance to post-production operations, with each person potentially holding multiple positions simultaneously. Meanwhile, because of the small team size, participants at every stage need to consider audience interaction and reactions, actively or passively participating in the "feedback-driven" audience interaction process. Due to this overlap and transformation of role boundaries in practice, participants can provide complementary insights into key issues such as workflows and feedback loops from different perspectives.
We also recruited micro-drama writers who have received traditional screenwriting training and worked as traditional screenwriters. First, the micro-drama creators community exhibits distinct "hybrid" characteristics, including both "transitional writers" who have shifted from the traditional film and television community and "native creators" without formal training backgrounds. Therefore, traditional screenwriters' experience serves as a crucial "baseline reference" to help understand how traditional narrative professional capabilities are reconstructed under data-driven, commercialization-oriented production models. Second, traditional screenwriters' "long-form video narrative logic" differs from micro-dramas' "iterative, short-cycle," audience-centered model. Third, traditional screenwriters represent "non-platformized creative labor," and their perspective can explain the changes in the creative ecology behind the creative transformation.

\subsection{Data analysis}
All themes and findings reported in this paper are derived from interview data rather than direct analysis of micro-drama content or platform data.
The primary material for data analysis is the interview data. The cases mentioned in the conclusion and discussion sections serve only as illustrative examples. In the actual data analysis of this study, we examined only the interview materials.We employed inductive thematic analysis. During the open coding phase, three researchers independently analyzed the data while collaboratively establishing an initial code set, then conducted preliminary coding exercises and discussed findings. Based on our RQs, we generated initial codes covering the micro-drama creation process (topic selection, episode division, writing) and audience participation methods (direct or indirect influence). Subsequently, through continuous code generation, theme categorization, and regular discussions to ensure reliability, we continuously reviewed and refined the codes as new data emerged, stopping data collection when core variables reached saturation.

\subsection{Ethical considerations}
This study strictly adhered to the anonymization agreement, institutional ethics review board guidelines, and the research protocol. All participants were clearly informed of the study objectives, data use, and privacy protections before providing informed consent. Participants retained the right to withdraw from the study at any time without penalty. After the interviews were completed, participants received appropriate compensation for their time and contribution to the research. All data collected in this study were anonymized, and all personally identifiable information was removed from the records. This ensures that there is no link between the collected data and the identity of each informant.

\subsection{Researcher positionality}
All members of the research team are micro-drama enthusiasts. One member has professional experience in the micro-drama industry, and one is a scholar with a background in traditional drama writing. The three research members have disciplinary backgrounds in communication studies, design, and film and television drama, respectively.The researchers' composition may introduce potential biases: as micro-drama enthusiasts, they may over-romanticize innovation while overlooking issues like content homogenization and over-commercialization; industry practitioners may lack critical reflection on structural problems such as inadequate writer protections and data-driven dependencies; scholars with traditional dramatic training may unconsciously apply classical narrative standards, underestimating micro-dramas' unique value as an emerging form; additionally, participants from specific regions may not represent diverse geographic and market practices, causing findings to disproportionately reflect particular platforms or areas. 

\section{Results}\label{sec:Results}

Based on the insights shared by our interviewees, this study first delineates the key differences between traditional film-production workflows and micro-drama creation, with particular attention to how creators adjust narrative structures and character development to accommodate the constraints and affordances of the vertical-screen format. We then examine the role of audience feedback and platform-generated data in driving content iteration. Although real-time incorporation of viewer comments has not yet become an industry-wide practice, some creators have begun integrating such feedback during continuous updates. Our coding framework is designed to uncover the underlying logic through which creators leverage audience input during the creative process, moving beyond surface-level observations to reveal how interaction and adaptation systematically shape the development of micro-drama narratives.

\subsection{The Evolution of Creative Workflow and Roles in Micro-Drama Production(RQ1)}\label{sec:Result}

\begin{figure*}[t]
  \centering
  \includegraphics[width=\textwidth]{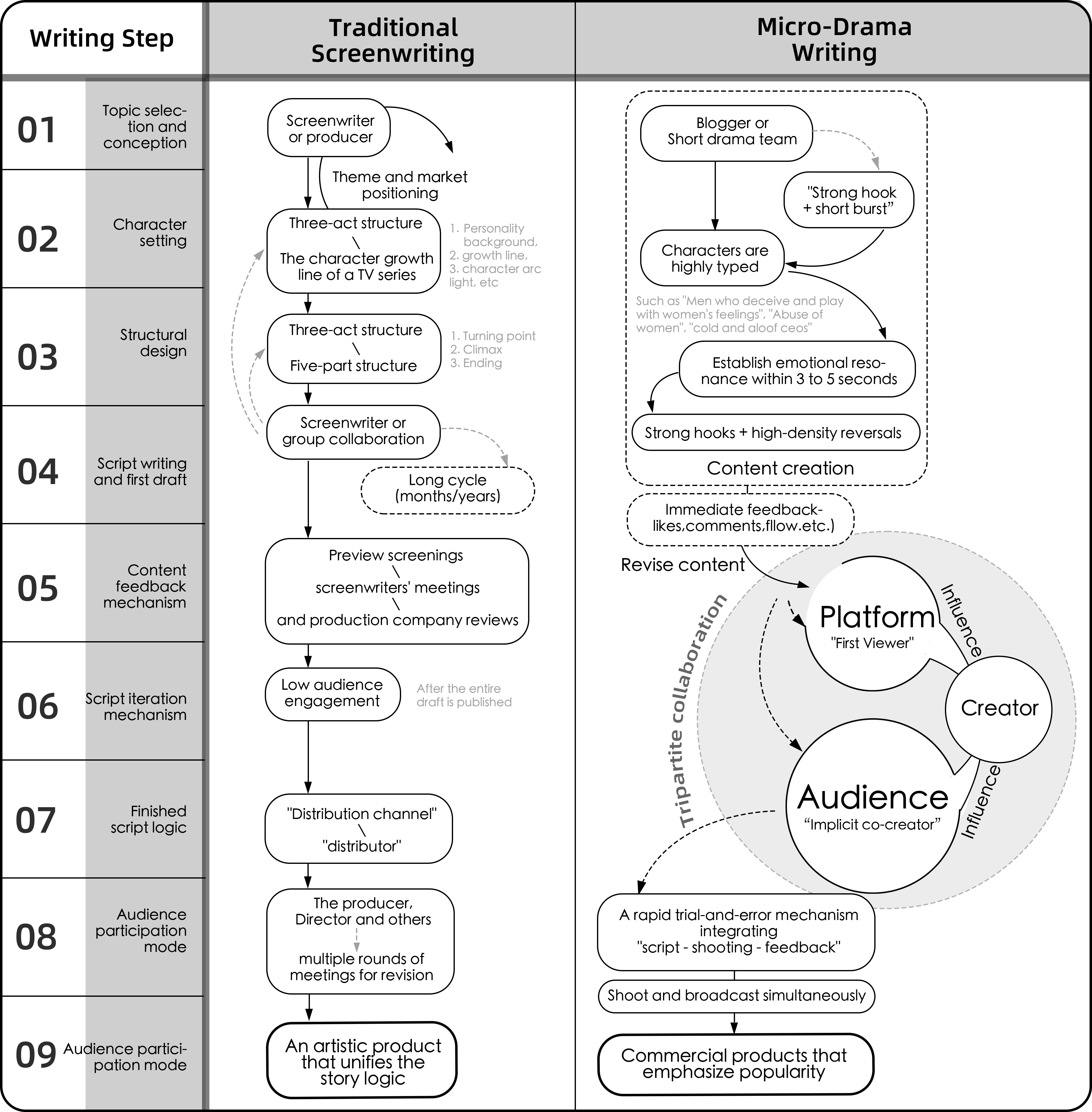}
  \caption{Comparative Workflow of Traditional Screenwriting and Microdrama Creation}
  \label{fig:workflow}
\end{figure*}

\textbf{4.1.1  From Linear Industrial Workflow to Platform-Driven Fast-Paced Production}
Compared to the traditional film and television production workflow illustrated in Figure \ref{fig:workflow}, micro-dramas are characterized by extremely compressed cycles, stronger market alignment, and markedly simplified creative logics. We observe systematic reconstructions in topic selection, script structure, and production models. In micro-drama production, the timeline from concept development to final release often spans only a few weeks or even days, with the entire process heavily shaped by platform trends and audience sentiment preferences.

During the interviews, participants consistently emphasized that topic selection is critical to a project's success and highlighted the clear differences between traditional film and television production and micro-drama creation. For long-form video projects, participants described topic selection as a ``process of long-term accumulation'' (P1, P8, P10, P16, P17). In contrast, for micro-dramas, they stressed the importance of following trending topics and adopting market-oriented strategies.  As micro-drama creator P3 noted, monitoring comment sections, popular memes, and audience interactions is the most effective way to identify topics with high resonance. Based on these signals, creators rapidly develop storylines that align with current audience interests in order to capture attention in a highly competitive environment. In practice, 
creators such as micro-drama director P8 and screenwriter P2 reported relying on trending lists, user comments, and meme culture on platforms such as Douyin, Kuaishou, and TikTok to identify high-frequency trends and emotional hotspots. These insights from creators who work across both traditional film production and micro-drama creation clearly demonstrate how platform logics exert a deep influence on creative decision-making from the earliest stages of production.

Screenwriters for television (P16, P17) and theatre (P18), all of whom have experience in both traditional film/TV and micro-drama production, noted that the two formats differ structurally in their narrative priorities. In traditional long-form drama, storytelling is predominantly character-driven, with an emphasis on character complexity and emotional arcs. As one television writer (P16) explained, ``An engaging plot must be built upon believable characters." In contrast, micro-dramas are widely described as “hook- and suspense-driven,” where narrative hooks take precedence over character depth. Several micro-drama creators (P10, P21, P24) summarized the typical structure as “a strong opening, rapid plot reversals, and a paywall to resolve suspense," relying on extreme scenarios and frequent twists to capture attention and sustain viewer engagement\cite{literature2024issue10}. Within this framework, characters are often simplified into binary oppositions, allowing audiences to quickly identify good and bad roles, reducing cognitive load, and eliciting immediate emotional responses. Although some creators emphasized creative freedom and noted that this model does not apply to all story types, the majority of interviewed micro-drama writers (P12, P13) acknowledged that their scripts are primarily organized around emotional stimulation , rather than character development as the driving force of the narrative.

Some participants noted that micro-drama creators commonly adopt a trial-and-error approach, producing content iteratively while shooting. This flexibility stands in stark contrast to horizontal long-form productions, which involve lengthy, highly linear workflows with limited room for adjustment. As one experienced television writer (P17) explained, ``Long-form content, from IP selection and adaptation to scriptwriting, filming permits, casting, and post-production, typically requires several years to complete. Each stage proceeds in strict sequence, leaving almost no space for on-the-fly adjustments."

In contrast, micro-drama production is characterized by flexibility and efficiency, with production cycles typically under one month, and some lightweight projects completed in just a few days. Two primary production models have emerged. The first model is the situational micro-drama format, in which episodes are released directly on user-generated content platforms such as Douyin, Kuaishou and TikTok. This format relies on real-time feedback, allowing creators to monitor audience preferences through comments and bullet screens and to adjust subsequent episodes accordingly. For example, after observing heightened viewer attention to a particular actor, micro-drama creator (P3) increased the actor’s close-up scenes in later episodes, forming a rapid cycle between audience input and creative decision-making; The second model is the platform-commissioned format, typically produced in collaboration with platforms such as TikTok, Tencent or iQiyi. Its workflow is more standardized. On-site editors are often involved to expedite revisions, and platforms systematically analyze viewer data such as keywords, completion rates and watch time. These analyses subsequently guide the development of storylines and character arrangements for the next production cycle.

Several micro-drama creators (P10, P12, P15) noted that the distribution mechanism of micro-drama platforms is tightly coupled with user preferences, where the performance of the first batch of released episodes often determines the trajectory of an entire series. This shift reflects a broader industry transition from users actively searching for content to content actively reaching users\cite{limingxia2024aesthetic}. Platforms construct user profiles by analyzing behavioral data such as viewing history, likes, and comments, and treat these high-frequency interactions as key signals of user stickiness and content value\cite{wangExploringActivitysharingResponse2024}, which then guide the recommendation of potentially relevant content. This mechanism is particularly prominent in the micro-drama ecosystem on WeChat Mini Programs. Micro-drama screenwriters (P2, P12) emphasized that such platforms provide almost no natural traffic, and content visibility primarily depends on recommendation algorithms and paid promotion. As a result, early performance metrics directly determine whether a series can continue to gain exposure, while underperforming content becomes algorithmically marginalized within a short period. A micro-drama director (P8) further explained that when initial views and interactions are insufficient, teams typically purchase targeted traffic and monitor data within a roughly 30-minute window to trigger secondary recommendations. The algorithm subsequently expands content visibility and audience reach based on this initial performance. To reduce investment risk, some micro-drama teams (such as P2, P4) release short clips before full publication to test audience reactions and adjust the content prior to official release. Participants viewed this “recommendation-driven feedback loop” as not only shaping content dissemination but also reshaping production workflows, making creative decision-making increasingly dependent on real-time data signals.

In terms of concrete production strategies, micro-drama screenwriters (P1) and micro-drama directors (P10, P12) commonly prefer reusing established narrative templates rather than experimenting with unconventional structures. Teams also optimize titles, cover images, and narrative pacing based on data indicators, such as employing keyword-dense expressions or intensifying emotional conflicts to boost click-through rates. For example, the micro-drama \textit{Shining Marriage} achieved 12.5 million views on the overseas platform FlixKeeper by amplifying title elements and character relationships. Under this distribution logic, creative teams routinely face constraints on narrative innovation.

\textbf{4.1.2  The integration of multiple job positions and the emergence of new occupations}
The traditional linear division of labor in the film and television industry has been disrupted, as micro-drama production has given rise to a creator-centered, multifunctional production model. This transition has also generated new occupational roles focused on platform-based data analytics and algorithmic recommendation.

\textbf{(1) Creators take on multiple roles, traditional job boundaries blur.} In traditional film and television teams, roles such as screenwriters, directors, and marketing personnel are clearly delineated, with well-defined responsibilities. In contrast, the micro-drama domain exhibits a markedly different division of labor due to its extremely short production cycles. Creators are often required to assume multiple roles, managing diverse tasks simultaneously to enhance efficiency and control costs. More importantly, the dissemination of micro-dramas is highly dependent on platform recommendation algorithms, which imposes multifaceted demands on creators. They must not only produce engaging content but also possess skills in data analysis and online platform management to align their work with algorithmic logic and user preferences. As micro-drama creators (P3 and P8) described, in order to maintain creative control within tight production schedules, they often handle the entire content production process themselves, from editing and scriptwriting to overall project planning. 
Even in AI-assisted micro-drama teams such as that of P9, the production groups are usually very small, typically consisting of only three to five members. Each individual is required to undertake multiple roles including directing, screenwriting, and editing in order to meet the tightly scheduled production pace and the efficiency demands of the industry.

\textbf{(2) Shaped by platform recommendation systems, the micro-drama industry has given rise to new professional roles such as traffic operation managers, content optimization specialists, and localization screenwriters}. These roles are primarily carried out within highly networked, distributed teams that rely on platform analytics dashboards, instant messaging tools, and collaborative editing software. Team members are responsible for core tasks such as designing advertising strategies, optimizing short-video re-edits, and adapting scripts across regions, thereby serving as critical links between content production and algorithmic recommendation systems. Among these roles, traffic operations managers play a particularly pivotal part during the release phase. As micro-drama writer (P1) summarized, reflecting an industry consensus, ``The script determines the minimum revenue, while advertising strategy determines the maximum revenue." Micro-drama senior editor(P12) noted that even when content is highly homogeneous, precise advertising placement can help a work overcome distribution bottlenecks and become a hit.

The advancement of cross-cultural production adds further complexity to this algorithm-driven workflow. Interviews revealed that in overseas markets like UK, US, and the Netherlands, micro-drama production teams often center around Chinese creators. Participants reported that machine translation exhibits clear limitations in producing natural language, resulting in time-consuming and fragmented script localization processes \cite{chen2025research,chenComparingNativeNonnative2025,xiaoSustainingHumanAgency2025}. Beyond language barriers, differences in social norms, figurative expression, political context, and geographic knowledge can introduce cultural discount effects \cite{singhPowerLanguageResisting2025}, weakening emotional resonance with local audiences. To address these challenges, micro-drama creators such as P4 and P11 hire local cultural consultants to study audience preferences and trends in depth, and collaborate with local writers and actors to adjust scripts, ensuring that content aligns with religious customs, linguistic habits, and value orientations in the target market. 

\textbf{4.1.3  From Consumers to Collaborators: The Reconfiguration of Creator–Audience Relationships}
Within the micro-drama platform ecosystem, the boundary between creators and audiences has become increasingly porous. Audiences can directly visit creators’ accounts to initiate conversations, and creator-generated content often appears alongside user-generated commentary. This produces a layered viewing experience \cite{huExploringDanmakuContent2024}, where interaction with creators and the possibility of receiving a response provide many viewers with social recognition \cite{kangStrangersYourPhone2016} and emotional support. At the same time, some users post provocative or extreme comments as a way to attract attention from creators or other viewers. The high accessibility of creator–audience interaction also generates benefits for creators. Even within loosely organized communities, creators can guide collective behavior through continuous engagement and resource sharing \cite{louSameMaskUnderstanding2025}. For example, creator P3 regularly replies to comments to enhance participation and strengthen community activity. Such interaction often creates a snowball effect. When creators respond, viewers tend to increase their participation and express appreciation for the attention, which encourages more users to join the discussion and fuels a positive cycle of engagement.

Community participation has thus become a central factor in retaining audiences in the micro-drama domain 
\cite{louSameMaskUnderstanding2025}. The role of creators extends beyond content publication; they actively adjust creative strategies based on platform data and community discussions. As micro-drama screenwriter P4 described, creators monitor metrics such as completion rates and content sharing to identify popular segments preferred by audiences and emphasize these in subsequent episodes. Some creators also establish dedicated WeChat groups or private chatrooms, providing paid users with early access to episodes and exclusive discussion spaces. Under this system, audiences assume a newly defined role. They are not only providers of feedback or test beds for narrative direction but also collaborators who contribute directly to narrative refinement and production decisions. This dynamic disrupts the traditional separation between production and consumption and highlights a collaborative narrative environment shaped jointly by platform mechanisms, community practices, and creator agency.

\subsection{Iterative Creation Mode Driven by Feedback Data from Audience (Q2)}\label{sec:Result}

\textbf{4.2.1 Platform Data and Viewer Feedback Inform Creative Decisions}

Micro-dramas have developed a distinctive ecosystem within the digital content industry, where data and audience feedback exert a profound influence on content iteration strategies. Understanding and effectively leveraging audience feedback is therefore essential \cite{brannonAudienceViewAIassistedInterpretation2024}, and they often dedicate substantial effort to studying audience needs and preferences \cite{whippleQualityQuantityPolicy2018}. Specifically, creators rely on two main types of information to evaluate and adjust content. The first type consists of quantitative operational metrics, including view counts, completion rates, likes, comments, and paid conversion rates. The second type comprises textual feedback collected from comments and bullet screens. 
Participants (P1, P5, P8, P10, P21) reported that these signals provide direct insight into audience perceptions. For example, creators reference trending micro-drama charts during early content planning to anticipate market trends, while platform metrics inform script evaluation. Micro-drama screenwriter P13 emphasized that revenue potential and conversion rates are primary factors in creative decisions. Beyond quantitative data, textual feedback offers emotional insights. Micro-drama creators (P5, P8, P10) actively monitor comments and live chat as authentic expressions of audience sentiment. Micro-drama director P8 noted that negative feedback is particularly informative for inferring audience profiles and latent needs, and writer P5 observed that live chat captures immediate emotional reactions.

Current practices lack systematic sentiment analysis tools, with most teams relying on rankings or third-party analytics. To address this gap, a few teams, such as P9’s, have begun developing bespoke AI tools to analyze traffic trends and audience profiles, providing data-driven support for content strategy.

\textbf{4.2.2 Feedback Mechanisms Permeate Every Stage of Production}
In the micro-drama industry, data and audience feedback have long surpassed their role as passive evaluation tools, evolving into a central control mechanism that actively shapes creative decisions. Content direction, production pacing, and monetization strategies are all deeply influenced by this feedback ecosystem.

\begin{figure*}[t]
    \centering
    \includegraphics[width=0.7\textwidth]{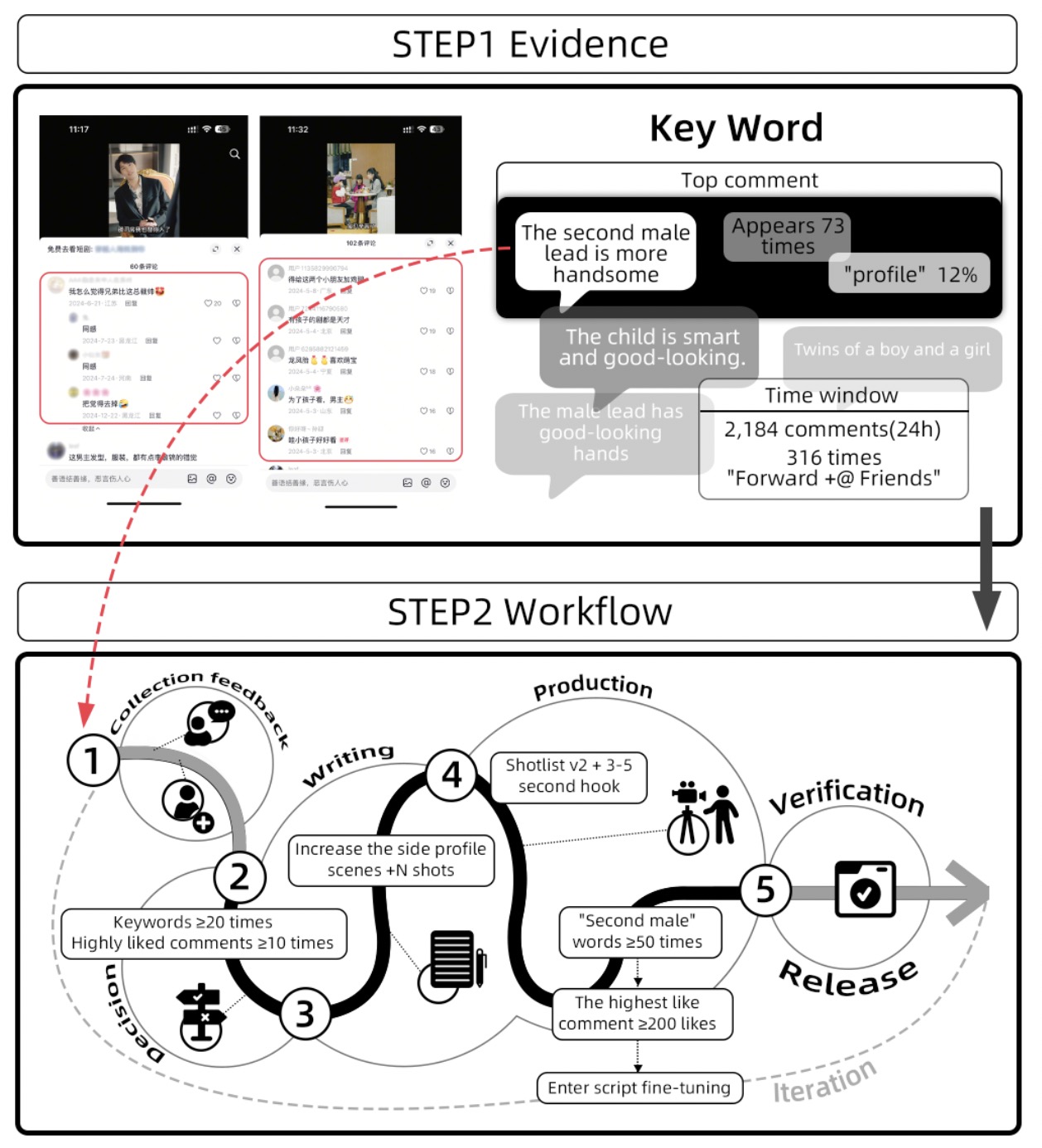}
    \caption{Evidence-backed feedback handling workflow: from audience comments to script changes and on-screen outcomes(Step1/2)
    Example source: ``Mr. Li, please sign for your twins, a boy and a girl"}
    \label{fig:Step1/2}
\end{figure*}

\begin{figure*}[t]
    \centering
    \includegraphics[width=0.7\textwidth]{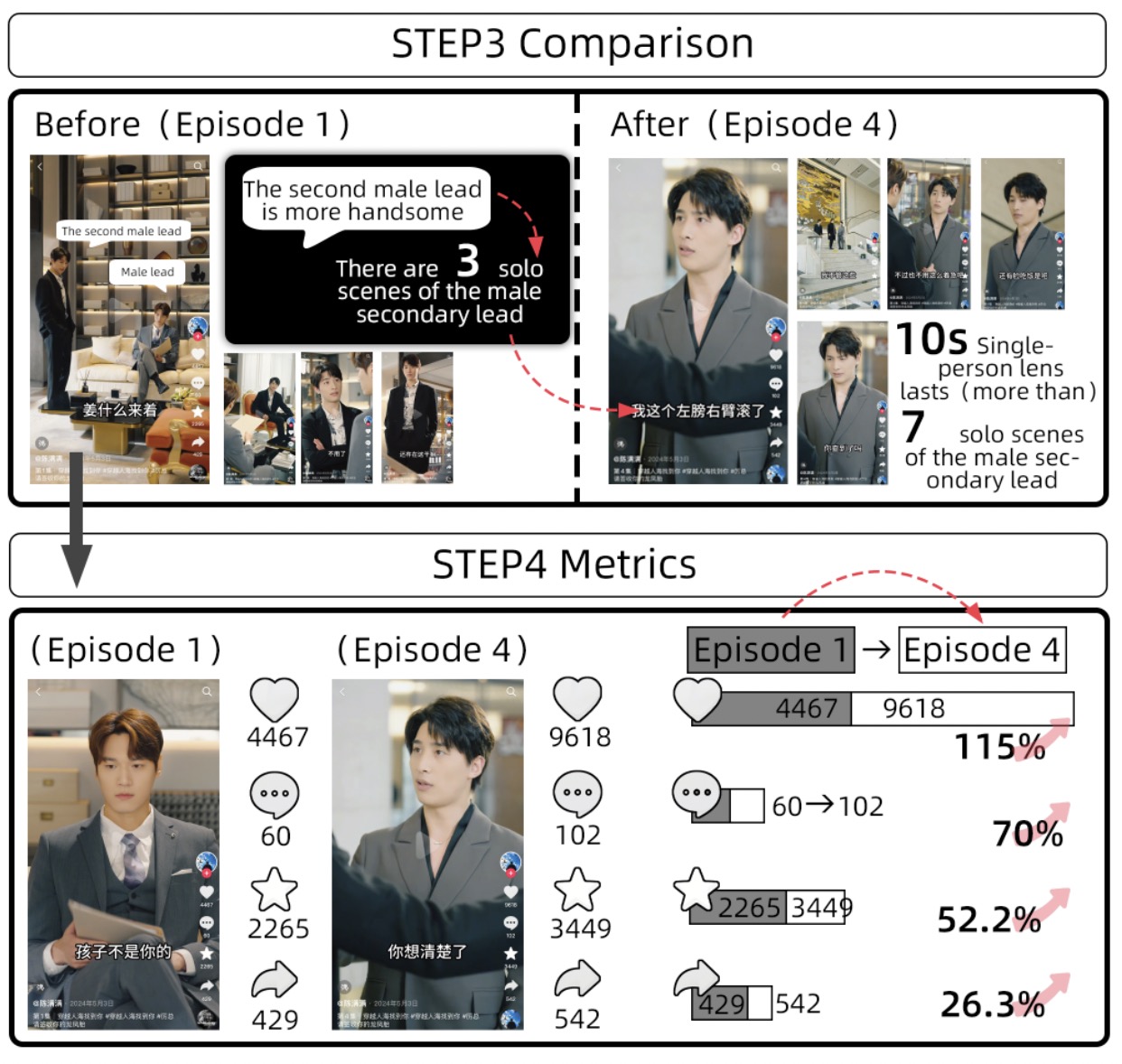}
    \caption{Evidence-backed feedback handling workflow: from audience comments to script changes and on-screen outcomes(Step3/4)
    Example source: ``Mr. Li, please sign for your twins, a boy and a girl"}
    \label{fig:Step3/4}
\end{figure*}

\textbf{In the early planning stage}, micro-drama creators rely heavily on data to guide topic selection and innovation. Micro-drama screenwriters (P1 and P5) monitor trending lists such as “system + CEO” or “rebirth + revenge” to identify high-conversion themes. Some micro-drama teams (P4 and P10), release “highlight test clips” to observe audience reactions to character pairing, costume design, and plot development, allowing early calibration. Drafts are refined through internal reviews and external editorial input (P1, P13). Independent creators such as micro-drama creator (P3) and director (P8) often use a flexible shoot and release approach to respond rapidly to audience interaction, while most companies prefer batch shooting with performance evaluated via platform analytics (P10). As shown in Figure \ref{fig:Step1/2} and \ref{fig:Step3/4} , creator P1 described a concrete case using the production \textit{Mr. Li, please sign for your twins, a boy and a girl}, initial audience comments repeatedly highlighted “second male lead is more handsome.” The team revised the script to increase this character’s scenes and close-ups, resulting in a 115\% increase in likes, 70\% increase in comments, and 263\% increase in shares by episode four. This demonstrates how keyword-driven feedback loops directly guide script and production adjustments. 

\textbf{In the post-production phase}, data continues to inform creative decisions. Teams hold regular analysis meetings to translate audience behavior and sentiment into new topics and narrative strategies (P13, P15). As micro-drama chief editor (P12) summarized, audience comments and platform metrics directly shape project decisions, determining whether a series is continued, pivoted, or expanded. Overall, micro-drama production has evolved into a “feedback loop–driven” model, where platform data and user feedback are central to creative strategy, requiring creators to balance traffic logic, monetization, and artistic expression.

\textbf{4.2.3 The Pitfalls of Overreliance on Platform Feedback}

\textbf{Questioning the authenticity and representativeness of data.}
The reliability of platform-provided data is frequently contested. High view counts or large volumes of comments are often driven by platform-level traffic boosting or paid promotional campaigns, rather than the intrinsic appeal of the micro-drama itself. Moreover, the voices reflected in comment sections are disproportionately shaped by a small group of highly active users, leaving the broader ``silent audience” underrepresented (P1, P2). Several participants even expressed concern that comment sections may include AI-generated interactions designed to artificially simulate engagement. As micro-drama director (P9) observed, \textit{"In today’s micro-drama market, AI-generated comments are commonly used to maintain the appearance of community activity. At some point, I stopped paying attention to comments, as it became impossible to discern between authentic audience responses and automated ones"}

\textbf{Limited access to meaningful feedback data.}Some small and medium production teams reported that their access to platform analytics was often restricted, forcing them to rely on third-party ranking websites or automated keyword summaries provided by the platform itself. However, such data are limited in scope and rarely provide actionable insights (P5, P7). An independent creator of micro-drama (P3) explained that \textit{ 'the micro-drama system generates a set of keyword summaries, which essentially groups popular terms into a few simplistic points, similar to what appears in trending topic summaries on Weibo. These summaries, however, hold little analytical value for guiding creative work” .} While some micro-dram teams (P1, P3, P9, P15, P25, P27) have begun experimenting with AI-based analytical tools, participants consistently noted that current AI systems still struggle with interpreting complex audience sentiments, such as irony, sarcasm, or nuanced criticism.

\textbf{Creative fatigue induced by overreliance on feedback.}Under the market logic that prioritizes rapid openings, frequent climaxes, and high conversion rates, many creators reported falling into cycles of meme-stacking and formulaic storytelling (P5, P15, P23). Such practices not only diminish narrative depth and originality but also accelerate content homogenization, leading to creative burnout among producers and aesthetic fatigue among audiences. As one micro-drama screenwriter explained (P26): ``Sustaining creativity in the current micro-drama ecosystem is inherently challenging. The industry requires creators to prioritize market demands over individual expression, and failure to do so can result in production teams being disbanded. While film still affords space for personal voice, micro-drama production is dominated by formulaic approaches that constrain innovation. This structural tension between market orientation and creative autonomy was a decisive factor in my decision to leave the company”.

\subsection{Emerging Styles and Thematic Innovations in Micro-Drama} \label{sec:Result}

Our findings, derived from interviews with creators and observations of their representative works, reveal that micro-dramas are not simply aesthetic updates of traditional audiovisual forms. Instead, they emerge as a new genre shaped by recommendation algorithms, audience feedback, and platform-driven market logics\cite{ryanNarrativeVirtualReality2015,phillipsJanetMurrayHamlet2017}. In this section, we first present how micro-dramas develop new thematic and stylistic directions, responding to RQ1. We then examine how creators adapt their workflows and modes of audience interaction to platform and cultural contexts, responding to RQ2.

\textbf{4.3.1 New Thematic Directions}

\textbf{(1) Everyday Social Conflicts.} Unlike the grand narratives of cinema, micro-dramas center on quotidian conflicts and socially resonant moments. Their ``domestication” of themes reflects not only creative preference but also alignment with algorithmic visibility. For instance, P8 emphasized: \textit{``Audiences want to see authentic emotions from daily life. Whenever I film something else, traffic declines.”}

\textbf{(2) Functional and Cross-Domain Narratives.} Beyond entertainment, micro-dramas increasingly integrate with commerce and tourism, creating hybrid ``content+function” formats. In e-commerce contexts, viewers purchase products directly while watching, blurring the line between storytelling and consumption. As one producer (P20) noted: \textit{``For many micro-drama companies, this is essentially an extension of live-stream commerce—at the end of the day, it is 
about selling goods.”} Similarly, collaborations with cultural and tourism bureaus use dramatized narratives to promote local heritage, extending micro-dramas beyond the entertainment domain.

\textbf{(3) AI-Driven Spectacle.} Generative AI has significantly lowered the threshold for genres traditionally requiring high budgets, such as fantasy, historical drama, and science fiction. This has enabled their rapid expansion in micro-dramas. An AI director (P8) emphasized: \textit{``Suspense and sci-fi fit AI the best, because they thrive on imagination, which is exactly where AI excels.”} These genres align with viewers’ appetite for spectacle in high-frequency, short-form consumption, while also reducing production costs.

\textbf{4.3.2 Reported Narrative Styles}

\textbf{(1) Fragmented-Integrated Storytelling.} Micro-dramas combine fragmentation with coherence. Unlike traditional series that rely on linear progression, micro-dramas adopt a ``micro-closure” structure—each episode delivers an immediate emotional climax while also sustaining an overarching plot. This duality supports fragmented viewing contexts (commutes, breaks) while fostering long-term engagement. Viewers oscillate between dipping into isolated segments and immersing themselves in extended storylines, reorganizing attention across both temporal scales.

\textbf{(2) Accelerated Industrial Rhythms.} The format’s vertical orientation and small frame intensify mobile-native habits. Plots often escalate rapidly, reaching dramatic turns within minutes. This industrialized rhythm corresponds both to fragmented attention and to algorithmic recommendation logics. Production has shifted from UGC spontaneity to PGC assembly lines, with studios and MCNs adopting serialized, industrial workflows. While this enables scaling, it also risks homogenization. Once a successful template emerges, it is quickly replicated, leading to what participants described as an ``unlikely creative crisis” \cite{linPlatformizationUnlikelyCreative2019}.

\textbf{(3) Interactivity and Participatory Narratives.} Participants noted that interactivity and participatory storytelling have become key characteristics of micro-dramas on social media platforms. This feedback loop allows narratives to evolve after release rather than remain fixed like traditional audiovisual works \cite{serimExplicatingImplicitInteraction2019}. Several participants described creating explicitly interactive micro-dramas. These works move beyond linear structures by using branching plots, networked narratives, and embodied decision points that actively involve viewers in story generation. Creators place suspense points, emotional peaks, and decision nodes to guide viewers to make story choices as characters. Such designs not only trigger emotional resonance at decision moments but also support the construction of viewers’ self-identity through role-play. Participants emphasized that these are not linear dramas with added interactions but rather a new storytelling practice shaped by platform logics and audience data.

In more conventional micro-dramas, comment text and emoji-based expressions also serve as feedback signals. Creators monitor these reactions to refine storylines, and audiences often extend stories into social media spaces through discussions, edits, and fan adaptations \cite{wangDLFEnhancingExplicitimplicit2025}. Some participants (P17) described this extratextual participation as part of a broader participatory storytelling practice.

Some independent creators (P3, P8) even respond to viewers in character. This role-based interaction blurs the boundary between fictional narratives and real communication, enhancing immersion and emotional engagement. Such practices are uncommon in traditional film or television production but are increasingly common in platform-based creative environments.

\textbf{(4) AI-Driven Micro-Dramas as a New Paradigm.} Interviewees described AI micro-dramas not merely as tool-assisted productions but as a human–AI co-creation paradigm. Our findings highlight three representative forms: AI-enhanced live-action, where tools support visual generation and audience analysis; anime-style micro-dramas, which transform novels or live-action content into animation while maintaining IP consistency; and hybrid productions that combine live-action with AI-generated environments and digital actors. These hybrid workflows not only diversify aesthetic possibilities but also reshape creator–audience expectations, demonstrating how technological affordances mediate human creative behavior.(P3, P4, P9)

\textbf{4.3.3 Cross-Cultural Adaptations}
Our further interviews indicate that Chinese companies exert strong influence in overseas micro-drama markets, supported by distinctive dissemination strategies. Platforms such as DramaBox, ShortMax, Goodshort, and ReelShort, founded and operated by Chinese firms, now reach audiences in over 240 countries.

We identify two parallel cross-cultural production strategies. The first is translation and dubbing. This low-cost, short-cycle approach is common in culturally proximate regions such as Southeast Asia, Japan, and South Korea, but it often suffers from a pronounced cultural discount in Western markets. As P4, a UK-based Chinese screenwriter, noted, some domestic genres (for example, reincarnation narratives) are less intelligible in Western contexts, reducing audience acceptance. The second strategy is localization through script rewriting and local production. This approach introduces locally familiar narrative elements and engages local actors, writers, and crews; it generally improves cultural fit but requires higher budgets and longer production cycles. These two strategies therefore create a tension between production efficiency and cultural adaptation, and platforms dynamically balance them according to commercial priorities.

Audience interaction patterns also vary cross-culturally. Our interviewee P14, a director working in the Netherlands, emphasized that Chinese audiences tend to prefer synchronous, collective engagement through danmu and comments, whereas Western audiences more often engage asynchronously and individually, for example via voting, branching choices, or community discussion on platforms such as Discord. A final observation is that repackaging and explanatory clips on TikTok can increase discoverability for original micro-dramas, but such content repackagers frequently lack professional translation skills and may weaken or alter contextual meaning. As a result, emphasizing ‘Eastern’ elements in titles, subtitles, or soundtracks can improve algorithmic visibility and attract overseas attention, yet it risks flattening cultural nuance. Platforms therefore make trade-offs among reach, cost, and cultural fidelity as part of their distribution strategies.

\section{Discussion}\label{sec:Discussion}
Our study investigates the iterative production practices of micro-dramas on Chinese social media platforms and the new forms of interactivity they afford for audiences \cite{seeringReconsideringSelfmoderationRole2020,veraTheyveOveremphasizedThat2025}. The findings suggest that, due to their pronounced market orientation and emotional bonding functions, micro-dramas have driven the evolution of the short-drama ecosystem into a new mode of storytelling—one where narratives are not pre-scripted in linear form but continuously reshaped through audience signals such as comments, memes, reposts, and trending hashtags. We situate these findings within the framework of participatory culture \cite{gomesConvergenceCultureWhere2008}, examining how micro-drama writers, audiences, and platforms jointly construct this feedback-driven mode of interactivity. In doing so, we highlight how micro-dramas transform the mechanisms of narrative generation and reconfigure the boundaries of interactivity in the social media era.

\subsection{Audience Feedback as an Engine for Rewriting Interactivity in Platform Narratives}
Transforming social media feedback into a narrative driver requires the coordinated participation of multiple actors. The reconfiguration of interactivity emerges through the co-creation of platforms, audiences, micro-drama writers, and broader sociotechnical environments \cite{plantinInfrastructureStudiesMeet2018}. While prior work on interactive storytelling often focuses on explicit interaction mechanisms such as branching pathways, pre-scripted trajectories, or viewer decisions \cite{riedlInteractiveNarrativeIntelligent2013,ryanNarrativeVirtualReality2015}, the rise of social media has reshaped the logic of interactivity. Short-video platforms now operate as platformized environments that structure both production and consumption through recommendation systems, visibility allocation, comment infrastructures, and community practices \cite{maityUnderstandingPopularitySocial2017,plantinInfrastructureStudiesMeet2018,quMicrobloggingMajorDisaster2011}.
Within such an environment, users become the primary actors served by and interacting with the platform. Zhou’s work \cite{zhouUsersRolePlatform2020} highlights that understanding users’ roles is essential because they are the core participants who shape and are shaped by platform-mediated content. Sun and Tang further argue that platforms can be optimized only when they genuinely reach and respond to broad user populations \cite{sunDouyinMyNourishment2025}. In this narrative environment, we want to emphasize that audiences are not passive viewers(4.2.1). Their comment keywords, memes, reposts, and collective reinterpretations accumulate into data traces that influence recommendation signals and ultimately reshape narrative pacing and contemporary media trends \cite{bulleyDualRoleStudent2024,lecompteItsViralStudy2021,maAmNotYouTuber2022}.

Based on interviews with micro-drama writers and observations of audience behavior, this study shows how feedback mechanisms permeate the entire production process(4.2.2/4.2.3). Writers do not respond only after release; audience signals such as emotional keywords, interpretive cues, and early sentiment are incorporated into story outlines, pacing decisions, casting, and the timing of plot reversals. Platform algorithms further amplify selected reactions, encouraging writers to convert these signals into narrative direction \cite{lecompteItsViralStudy2021,bulleyDualRoleStudent2024}. In this process, feedback operates as a distributed interactional system. Audiences participate not through discrete choices but through continuous behavioral traces that shape production rhythms. Prior work shows similar dynamics, where writer–audience interaction emerges from ongoing feedback flows rather than isolated inputs \cite{choiProxonaSupportingCreators2025}. The low-friction interfaces of short-video platforms intensify this process by enabling immediate responses and rapid creative iteration, which accelerates narrative adjustment.Our findings also align with sociotechnical research on creative labor that describes platform storytelling as a negotiation among writers, audiences, and algorithmic infrastructures \cite{simpsonRethinkingCreativeLabor2023}. In micro-dramas, this negotiation becomes visible as platforms curate audience signals, writers adapt to them, and audiences continue to reshape the data that guides production. These dynamics form a dynamic narrative community in which interactivity develops through sustained feedback cycles rather than predefined mechanisms.

Platform recommendation systems help viewers quickly locate content that aligns with their interests, and multi-layer interaction mechanisms provide accessible social channels through which users experience being noticed, recognized, and acknowledged online \cite{sunDouyinMyNourishment2025}. Sun and Tang show that such mechanisms strengthen users’ sense of presence and identity within social media environments, which encourages more active participation in narrative platforms \cite{sunDouyinMyNourishment2025}.Our findings highlight another dimension. This seemingly natural participation can also reflect coordinated actions by platforms and writers who depend on continuous streams of feedback. Platforms may surface highly personalized content that aligns with users’ preferences or aversions to stimulate emotional fluctuation and sustain engagement \cite{wanHashtagReappropriationAudience2025}. Writers may similarly deploy exaggerated or stereotypical characters to provoke reactions and amplify visibility. These dynamics reveal the risks of overreliance on platform feedback. Real-time responses can guide writers toward resonant themes, yet they also narrow creative space by prioritizing majority preferences. Rapid iteration encourages writers to optimize for exposure metrics rather than narrative exploration, reinforcing formulaic patterns \cite{jonasDigitalStorytellingDeveloping2022}. Some feedback signals are intentionally shaped by platforms or writers, limiting their value for meaningful narrative co-creation. In addition, the volatility of audience emotions introduces further uncertainty, particularly when interpretations differ across social backgrounds, cultural resources, or fan communities \cite{yousefiExaminingMultimodelEmotion2024}. These patterns show an uneven distribution of agency within feedback-driven ecosystems. Audience signals influence narrative directions, but platform infrastructures ultimately determine which signals are amplified and how they shape the production process \cite{sunDouyinMyNourishment2025}.

These findings show that platforms shape users’ judgments and feedback practices in meaningful ways. Yet, unlike environments such as WeChat where users can customize functions through official accounts or mini-programs \cite{zhouUsersRolePlatform2020}, micro-drama audiences have limited influence on platform functionality. Their participation typically appears as lightweight signals such as comments, emojis, and reposts rather than deeper forms of customization or structural change.This limitation can be explained by power asymmetries and gaps in digital literacy. Writers rely on platform infrastructures due to unequal access to resources \cite{maAmNotYouTuber2022}, and audiences often lack the technical capacity to intervene in platform features \cite{maityUnderstandingPopularitySocial2017,quMicrobloggingMajorDisaster2011}. Even so, some audience-driven customization is mediated through writers adaptation and platform governance. Moderators may amplify recurring keywords or emotional cues, which then guide later narrative adjustments. This suggests that feedback-driven interactivity is not simply the result of audience voices, but emerges from ongoing negotiations among audiences, micro-drama writers, and the platform infrastructures that mediate their interactions.

Although platform recommendations and writers’ production routines are continuously refined through audience feedback, resulting in experiences that feel increasingly personalized, this dynamic presents both benefits and risks for viewers. On the one hand, personalization allows audiences to locate preferred content more quickly. On the other, it introduces drawbacks such as search inertia and limited protections for personal data privacy \cite{veraTheyveOveremphasizedThat2025}.Our study further identifies concerns related to aesthetics and identity formation. Real-time feedback in platform-mediated environments accelerates iteration and helps writers align with audience expectations, but it can also constrain audiences within familiar interpretive frames, reducing their exposure to new perspectives or narrative possibilities \cite{veraTheyveOveremphasizedThat2025}. In addition, viewers rarely express concern about privacy or data use, including information such as personal details or location. Many regard their comments as spontaneous acts of self-expression, although these traces may be shaped by platform incentives, writers’ strategies, and limited awareness of privacy risks \cite{gengHowSocialMedia2024}(4.1.3). Observations from micro-drama comment sections reveal that viewers often share sensitive personal information, which may be captured and used by both platforms and writers.

These findings indicate that feedback-driven interactivity carries both productive and problematic implications. While real-time responses support rapid narrative adjustment and closer alignment with audience expectations, they also narrow narrative diversity and draw users into opaque systems of data capture. Understanding and reshaping the interactions among platforms, micro-drama writers, and audiences therefore requires careful attention to the trade-offs within feedback mechanisms, including both their enabling functions and their vulnerabilities. We argue that building more transparent, equitable, and sustainable forms of collaborative storytelling will depend on coordinated efforts among all stakeholders within platformized media environments.

\subsection{Expanding the Boundaries of Micro-Drama Writing}
Compared with traditional scriptwriting scholarship, academic discussions have primarily focused on the formalization of narrative structures, character development, and collaborative workflows. Classical works emphasize paradigms such as the three-act structure \cite{fieldScreenplayFoundationsScreenwriting2005}, character arcs \cite{campbellHeroThousandFaces2004,chen2025research}, and the teamwork of script development \cite{zhangFrictionDecipheringWriting2025,mongeroffarelloInvestigatingHowComputer2025}.
As a rapidly expanding form of digital storytelling, micro-dramas fundamentally challenge these theoretical foundations and compel a reconsideration of the very boundaries and meanings of “writing.” Echoing McLuhan’s proposition that media extend human perception and agency \cite{fen1969marshall}, the rise of micro-dramas illustrates how such extensions reshape creative writing practices. This shift expands the social boundaries of writing (through audience–platform–micro drama writer co-configuration), its media boundaries (via multimodal expression), and its methodological boundaries (from experience-driven to data-sensitive processes). 

One core finding of our study is that micro-drama audiences do not act as “co-writers” in the way short-video users often do. Instead, they participate through comments, affective reactions, and behavioral data, indirectly steering narrative decisions. We conceptualize this as narrative steering participation, driven by interpretive rather than productive acts. Due to the serialized nature of micro-dramas and the logic of algorithmic distribution, such participation confers a form of temporal narrative influence not found in short-video environments. Metrics such as completion rates, drop-off points, and discussion intensity not only shape distribution but also retroactively inform pacing, character allocation, and conflict sequencing, pushing micro-drama writing toward a dynamic, iterative, and data-driven process. Writers frequently adjust plot direction based on real-time audience signals, exposing new boundaries of authorship under platform conditions: micro-drama writers must continually negotiate between responding to audience preferences and maintaining narrative integrity, and between data-driven optimization and preserving authorial judgment. This aligns with prior work on creative agency \cite{tafesse2023content,hamiltonStreamingTwitchFostering2014}, platform power \cite{bhargavaCreatorEconomyManaging2022,devitoPlatformsPeoplePerception2017}, and algorithmically mediated labor \cite{simpsonRethinkingCreativeLabor2023,bonifacioFansRelationalLabor2023,duffy2017not}. We argue that micro-dramas instantiate an algorithmically augmented participatory narrative model distinct from both short-video interaction and traditional audiovisual production: audiences do not produce textual content, yet their feedback—amplified by algorithmic ranking and accumulated across rapid update cycles—exerts substantial influence over narrative direction.

We also find that AI tools have become widely integrated into micro-drama writing workflows (see Section 4.3.1), giving rise to what writers describe as an emerging paradigm of AI-augmented micro-drama writing. Although writers reject the feasibility of LLMs independently producing publishable scripts—citing limitations in dialogue nuance, emotional texture, and structural coherence (P3, P9, P17), consistent with observations by Grigis \cite{grigisPlaywritingLargeLanguage2024}, they nonetheless acknowledge the value of AI for character modeling, plot scaffolding, and creative ideation. These perspectives resonate with Tang et al.’s \cite{tangUnderstandingScreenwritersPractices2025} proposals for enhancing narrative reasoning through graph-based emotion modeling and visual interfaces for understanding narrative space. Yet, AI adoption simultaneously raises new boundary questions concerning narrative control, originality, and copyright \cite{limaPublicOpinionsCopyright2025,heWhichContributionsDeserve2025}. Based on our findings, we argue that future writing tools should enhance creative capacity without undermining writers’ narrative judgment and agency.

Taken together, our findings suggest that the evolution of micro-drama writing reconfigures both the technological and social scaffolding of creative practice. Future research and tool development should consider not only how algorithms and AI can support narrative design, but also how they shape authorship, audience influence, and creative boundaries.

\subsection{Design Implications}
This study offers three design implications for the design of interactive systems and platforms that support micro-drama feedback-driven narratives.

Short-video platforms and micro-drama production practices have indeed expanded contemporary narrative environments \cite{duUnderstandingEffectOpinion2025}. By embedding audiences into ongoing feedback cycles, these platforms strengthen viewers’ sense of participation and presence while offering creators an unprecedented large-scale “real-time testing ground” that shapes emerging modes of storytelling and creative practice \cite{veraTheyveOveremphasizedThat2025}. Yet our findings show that this feedback-driven mode of creation is accompanied by clear limitations. From the audience perspective, feedback is easily guided by platform logics and writer strategies, influencing aesthetic boundaries and identity formation. From the writer’s perspective, continuous pressure from audience signals and algorithmic expectations often compresses narrative diversity and creative autonomy. From the platform side, current systems focus primarily on capturing and amplifying feedback rather than treating feedback as something that can be configured, re-structured, or governed. Based on these observations, we outline three directions for design implications.

\textbf{Designed as a configurable feedback system.}
From the platform perspective, our findings indicate that feedback on short-video platforms is often treated as a naturally occurring data stream rather than as a configurable and interpretable design material. Prior work shows that user feedback is shaped by platform infrastructures and moderation practices, which implicitly define what becomes visible and meaningful in the interaction environment (e.g., Seering’s work on how platforms structure participation and guide user responses) \cite{duUnderstandingEffectOpinion2025}. This suggests an opportunity to reconceptualize feedback not simply as data to be captured but as an adjustable environment that can be configured to support different creative and interpretive processes. If platforms were designed to provide writers with multiple “feedback views,” writers would be able to explore alternative narrative conditions before responding to audience reactions. For example, what would happen if early-stage commenters—rather than the full audience—were allowed to influence the emerging narrative arc? How might weighted feedback based on audience diversity rather than sheer volume shift character development or pacing? 

At the same time, Shen’s study shows that platform decisions often depend not on content alone but on “ambient elements”—the types of users a video attracts and the interactions surrounding it \cite{veraTheyveOveremphasizedThat2025}. This insight raises an important design tension: if configurable mechanisms let writers balance different audience compositions, can platforms also redistribute attention in ways that mitigate structural bias against minority perspectives?

The system built around these configurable views can enable researchers and platform teams to examine how different feedback environments shape narrative diversity and creative decision-making. A common approach in interaction research is to use controlled A/B testing to compare how variations in interface presentation influence micro-drama writers’ practices, as demonstrated in Suh et al.’s experimental study on creative platform interfaces \cite{veraTheyveOveremphasizedThat2025}. In this context, contrasting a “default feedback view” with a “configurable feedback view” offers a way to evaluate their effects on story structure diversity, character complexity, and writers’ perceived autonomy.

\textbf{Designing co-creation feedback systems with filtering affordance.}
From the perspective of writers, one central challenge is the intensity of feedback cycles. When signals arrive too quickly or too frequently, they generate information burden and narrative pressure, making it difficult for writers to sustain a sense of creative autonomy. This highlights the need for systems that offer protective mechanisms which help writers manage feedback rather than be driven by it.

Prior work on writing support systems demonstrates that adjustable feedback timing can preserve writers’ focus. For instance, Carrera et al.’s Nabokov's Cards system \cite{carreraNabokovsCardsAI2025} allows writers to regulate when assistance appears, enabling them to enter a concentrated writing mode without continuous external prompts. Similar insights appear in Weber et al.’s LegalWriter \cite{weberLegalWriterIntelligentWriting2024}, which provides structured layers of revision so writers can selectively engage with feedback at appropriate stages.

Building on these ideas, platforms supporting micro-drama creation could incorporate filtering modes that temporarily delay comments, hide volatile metrics, or allow writers to “lock in” narrative segments before re-entering the broader feedback environment \cite{carreraNabokovsCardsAI2025,weberLegalWriterIntelligentWriting2024}. Such mechanisms support writers in a technologically mediated production context by enabling more intentional integration of audience signals rather than reactive adjustments under high-frequency feedback.

\textbf{Designing Co-Creative Feedback Ecologies that Enable Identity Diversity.}
From the audience perspective, one direction is to shift the emphasis from short-term signals such as completion rates or rapid clicks to longer-term satisfaction metrics that better reflect sustained engagement \cite{yousefiExaminingMultimodelEmotion2024}. Another is to incorporate mechanisms that safeguard minority perspectives, ensuring that dominant audience preferences do not marginalize underrepresented narrative voices \cite{augustKnowYourAudience2024,veraTheyveOveremphasizedThat2025}. Systems can also encourage viewers to encounter a broader range of character arcs and narrative structures rather than being continually funneled back into pre-existing interest profiles \cite{veraTheyveOveremphasizedThat2025}. Finally, attention to privacy boundaries remains essential, particularly as audience expressions often include sensitive personal information that can be inadvertently captured and repurposed \cite{gengHowSocialMedia2024}.

A promising direction for future work is to build an experimental platform that integrates configurable feedback views, diversity-oriented ranking mechanisms, generative AI features, and protective control features for creators \cite{lingSketcharSupportingCharacter2024,zhangPickRightThing2025}. Prior work has shown the value of such testbeds. Jiang et al. used an experimental Douyin-like interface to study how platform cues shape user boundary negotiation \cite{gengHowSocialMedia2024}, and VoxPop demonstrated how alternative interaction infrastructures can shift participation and discourse \cite{sharevskiVoxPopExperimentalSocial2021}. Inspired by these methods, a prototype micro-drama platform could manipulate feedback configurations in a controlled setting to generate empirical evidence of how different design interventions influence narrative diversity, character representation, and audience engagement.

\subsection{Limitations}
\textbf{5.4.1  Self-Reported Narrative Styles.}
The narrative styles and thematic innovations identified in our study are primarily based on micro-drama writers’ self-reports rather than systematic validation against the broader corpus of micro-dramas on social media platforms. While participants frequently described emerging styles they experimented with, such as condensed emotional climaxes or cross-role performances, these accounts were largely retrospective and impressionistic. Without triangulation using large-scale content analysis or platform-level data, our findings may overrepresent idiosyncratic practices. Future research should integrate computational methods, including corpus-based stylistic analysis or longitudinal tracking of serialized works, to corroborate micro-drama writers’ reported innovations and provide a more representative mapping of stylistic trends.

\textbf{5.4.2  Subjectivity in Writers’ Accounts.}
Closely related, micro-drama writers’ descriptions of how audience feedback shaped their work were highly subjective. Participants often relied on ``feeling” the audience’s expectations or noticing repeated comments, but they seldom articulated concrete, traceable workflows showing how feedback translated into revisions of scripts, performances, or edits. This reliance on memory and interpretation introduces potential bias, as writers’ perceptions of audience intent may not align with actual audience responses or platform metrics. To address this limitation, future research could triangulate writers’ narratives with direct audience interviews, comment analysis, or platform-provided engagement data, offering a more precise account of how feedback loops operate in practice.

\textbf{5.4.3 Limited Consideration of Audience Perspectives.}
Our study primarily foregrounds the perspectives of micro-drama writers, which means the role of audiences was only indirectly addressed. While participants often reflected on audience reactions, these accounts were filtered through the writers’ interpretations rather than derived from viewers’ own voices. This absence of direct audience perspectives limits our ability to fully capture the dynamics of feedback loops, because audience practices such as commenting, meme-making, or sharing are socially situated activities that may carry meanings different from those perceived by micro-drama writers. Without systematically incorporating audience experiences, our analysis risks overlooking the ways in which viewers themselves negotiate attention, emotion, and participation, all of which shape the evolution of micro-drama narratives. Future research should therefore directly engage with audience perspectives through interviews, ethnographic studies of fan practices, or analysis of large-scale comment data, in order to complement writers-centered accounts and build a more comprehensive understanding of the ecosystem.

\textbf{5.4.4  Lack of Cross-Cultural Comparative Analysis.}
Our study primarily focused on Chinese micro-drama writers, leaving unexplored how micro-drama practices may vary across cultural and regulatory contexts. While the Chinese market currently dominates global micro-drama production, the international expansion of the format raises important comparative questions. Prior work on platform studies has emphasized that content production is deeply shaped by socio-technical environments and regulatory regimes, leading to divergent practices across regions \cite{burgess2018youtube,plantinInfrastructureStudiesMeet2018}. For instance, Western micro-drama writers working on TikTok often operate under different industry norms, monetization models, and content regulations than their Chinese counterparts \cite{choudhuryGenderCrossculturalDifferences2017,livingstoneReframingMediaEffects2016,wangExaminingAmericanChinese2016}. A micro-drama writer in the U.S. might prioritize alignment with brand sponsorship or copyright restrictions, whereas Chinese micro-dramatists are more attuned to rapid audience validation and platform traffic incentives. Likewise, cultural expectations around themes such as romance, family values, or political sensitivity differ substantially between contexts, potentially leading to divergent narrative strategies \cite{sunBridgingCulturalDifferences2023}. The absence of such perspectives limits the generalizability of our findings. Future research should therefore incorporate cross-cultural comparisons to illuminate both the convergences and differences in feedback-driven creative practices across regions.

\textbf{5.4.5  Methodological Constraints.}
A methodological limitation of this study concerns the scope and composition of our data. Our analysis relied primarily on semi-structured interviews, which offered rich accounts of writers’ perceptions and decision-making processes but did not allow us to systematically examine how these practices appear within the texts and aesthetics of the micro-dramas themselves. Although participants described how audience feedback shaped plots, pacing, and character choices, we did not validate these accounts through direct comparisons with the narrative structures, stylistic patterns, or episode-level developments present in their published works. Prior HCI and media studies have shown that integrating interviews with large-scale content analysis or computational text methods can reveal how creative practices materialize at the level of narrative form and genre evolution \cite{burgess2018youtube,wohnAudienceManagementPractices2020}. Without such triangulation, our findings remain grounded in self-reported experiences and cannot fully substantiate claims about emerging storytelling conventions.

A second constraint relates to potential sampling bias. As described in our method section, all participants were based in China and worked within the highly fluid role structures characteristic of micro-drama production. These writers often shift between planning, writing, acting, editing, and audience engagement, which provides holistic insight into feedback-driven creation but may obscure differences between more specialized or industrialized production environments. Moreover, our sample included both “native” micro-drama writers and traditional screenwriters who entered the short-form video industry. While traditional screenwriters served as an important baseline for understanding shifts from long-form narrative logic to iterative, rapid-cycle storytelling, their perspectives may not represent writers without formal training or those operating within larger commercial teams. These characteristics limit the diversity of creative strategies and collaborative models reflected in our dataset.

Future work should therefore expand this methodological framework by combining interview data with corpus-based approaches, such as episode-level comparisons, computational text analysis, and multi-modal annotation of representative micro-dramas. Such integration would allow researchers to trace how feedback loops appear not only in micro-drama writers’ reflections but also in the evolving narrative forms, character trajectories, and stylistic signatures present in the content itself. Broader sampling across regions, micro-drama writers types, and production structures would further clarify how feedback-driven storytelling varies across platform cultures and creative ecosystems.

\section{Conclusion}\label{sec:Conclusion}
In conclusion, this study examined the feedback-driven production practices of micro-dramas on Chinese social media platforms. Drawing on interviews with 28 creators, micro-drama writers, and actors, along with analyses of representative works, we showed how implicit audience signals—such as comments, memes, and reposts—reshape narrative logics. Our findings challenge the author-centered paradigm in scriptwriting scholarship, reconceptualize writing as a socio-technical practice embedded in continuous feedback loops, and highlight the triadic negotiation among micro-drama writers, audiences, and platforms.

As participatory culture plays an increasingly central role in scriptwriting, we argue that future design should focus on developing tools and infrastructures that enhance the transparency, inclusiveness, and creative autonomy of feedback cycles. Such efforts can balance narrative diversity, market imperatives, and user influence, contributing to a more open and responsible ecosystem of feedback-driven storytelling.




\newpage
\newpage
\newpage
\appendix

\label{sec:Appendix}

\section{Interview design and question framework}
\subsection{Research Question 1: What emerging roles and production workflows have writers adopted in production of micro-dramas?}

\subsubsection{Basic Information and Role Positioning}
\begin{itemize}
    \item Basic information (age, education, years in industry, current platform)
    \item What are your main video creation types and your role division? (screenwriter/director/video creator/MCN operator)
    \item What are the essential differences in creative workflow between "micro-drama writers" and traditional "screenwriters"? Have you received professional screenwriting training?
\end{itemize}

\subsubsection{Creative Process and Team Division}
\begin{itemize}
    \item How does the screenwriting team divide work?
    \item What is the typical process for creating a micro-drama? (concept, outline, episode breakdown, draft, revision, finalization, etc.)
    \item How is feedback communicated? What are the fixed procedures from receiving feedback to implementing script revisions?
\end{itemize}

\subsubsection{Creative Techniques and Strategies}
\begin{itemize}
    \item What unique techniques do you use for story pacing control? Are there "rules" like "golden three seconds" or conflict within 15 seconds?
    \item How do you collect and filter viral elements? What methods or channels do you use to track popular formulas?
    \item Do you have "viral formulas" or modify successful cases to create new works?
\end{itemize}

\subsubsection{IP Adaptation vs. Original Scripts}
\begin{itemize}
    \item What are the differences in decision-making mechanisms between IP adaptation and original scripts at the creative initiation stage?
    \item In IP adaptation projects, what role do original novel data and reader feedback play? Which has greater weight compared to micro-drama market feedback?
\end{itemize}

\subsubsection{Professional Feedback Mechanisms}
\begin{itemize}
    \item What feedback do you typically receive from professionals? At which stage does it intervene?
    \item What aspects does professional feedback focus on?
    \item When editors reject artistic designs citing "user preference data," do they provide specific data support?
    \item How do professional feedback and audience feedback differ in focus and importance? How do you integrate both?
\end{itemize}
\subsubsection{AI Tool Applications}
\begin{itemize}
    \item Have you tried using AI tools to analyze audience comments or market data?
    \item Do you use AI to assist script creation? In which stages does AI provide effective help?
\end{itemize}

\subsection{Research Question 2: How do writers work with audiences to iteratively shape the plots and characters for micro-dramas?}

\subsubsection{Feedback Sources and Weight}
\begin{itemize}
    \item Among audience, market data, and professional feedback, which has the greatest influence? Why? Does this change across project stages?
    \item Through which platforms and channels do you obtain audience feedback? Which platform is most valuable?
\end{itemize}

\subsubsection{Feedback Collection and Filtering Mechanisms}
\begin{itemize}
    \item How do you collect feedback? Through technology or experience? How does the backend operation work?
    \item How do you filter and process massive feedback? Which types of feedback attract your attention more easily?
    \item How do you filter information related to "plot progression and character development"? Do you have specific criteria?
    \item Does your company have a comprehensive feedback collection and analysis system? How does it operate?
\end{itemize}

\subsubsection{Feedback Impact at Different Stages}
\begin{itemize}
    \item Do you closely monitor real-time audience feedback? Throughout or at specific stages?
    \item At the initial creative stage, which factors have the greatest impact?
    \item In early creation, do you reference audience feedback from existing works to decide your topic or character settings?
    \item Do you release preview clips to test audience reactions? Do reactions change your story framework?
    \item Can real-time audience feedback produce "real-time intervention" on narrative pacing and plot structure?
\end{itemize}

\subsubsection{Market Data and Evaluation Indicators}
\begin{itemize}
    \item How do you quantitatively evaluate script success? Which market feedback indicators are most critical?
    \item How long does it take to obtain revenue data? How do you use it to adjust subsequent creation?
    \item Does short video "re-creation" content impact your subsequent creation?
\end{itemize}

\subsubsection{Feedback's Impact on Creation}
\begin{itemize}
    \item Does audience feedback help you "correct mistakes" or "re-understand characters/stories"?
    \item Have you adopted audience suggestions from comment discussions about plot development? Have audiences partially "taken over the story"?
\end{itemize}

\subsubsection{Theoretical Reflection on Author-Audience Relationship}
\begin{itemize}
    \item With audience feedback driving narrative iteration, how has the "author's" creative agency changed?
    \item Do you feel you and the audience are "co-creators"? When do you most feel this interaction? Do audiences seem more like reviewers, collaborators, or update-demanding machines?
    \item Is micro-drama no longer "an author telling a story" but "audiences deciding the story together"?
    \item How does audiences becoming "co-creators" impact screenwriter professional identity? Are there risks of "narrative losing control due to over-catering to feedback"?
\end{itemize}

\subsubsection{Ideal Feedback Mechanism}
\begin{itemize}
    \item What should an "ideal" feedback mechanism conducive to healthy micro-drama creation development look like?
\end{itemize}
\section{Platform introduction}
Currently, users primarily watch micro-dramas through specialized micro-drama platforms, short-video platforms, and long-form video platforms. The following sections introduce micro-drama-related platforms from the perspectives of platform type, platform format, and content creation model.(The following are newly added tables providing detailed introductions to different short drama platforms.)

\begin{table*}[t]
\centering
\caption{Specialized Micro-Drama Platforms}
\label{tab:specialized_platforms}
\begin{tabular}{@{}ll@{}}
\toprule
\textbf{Category} & \textbf{Description} \\
\midrule
Platform Name & Hongguo / Hema Theater \\
& \\
Platform Overview & \begin{tabular}[t]{@{}p{10cm}@{}}
Specialized micro-drama platforms operating on a free-to-watch \\
model with ad-supported monetization. They offer micro-dramas \\
across urban, fantasy, and romance genres, plus film and novel \\
sections. Viewers unlock episodes by watching advertisements.
\end{tabular} \\
& \\
Viewer Interaction & \begin{tabular}[t]{@{}p{10cm}@{}}
Viewers earn small cash rewards for watching content. Platforms \\
encourage comments and likes.
\end{tabular} \\
& \\
Content Creation & \begin{tabular}[t]{@{}p{10cm}@{}}
Platforms acquire content through producer submissions or direct \\
purchases. Rankings and metrics display trending micro-dramas.
\end{tabular} \\
\bottomrule
\end{tabular}
\end{table*}

\begin{table*}[t]
\centering
\caption{Short-Video Platforms}
\label{tab:short_video_platforms}
\begin{tabular}{@{}ll@{}}
\toprule
\textbf{Category} & \textbf{Description} \\
\midrule
Platform Name & Douyin / Kuaishou/instagram/.Tiktok \\
& \\
Platform Overview & \begin{tabular}[t]{@{}p{10cm}@{}}
Decentralized short-video social media platforms where micro-dramas \\
are not the primary revenue source. Built-in mini-programs allow \\
users to access micro-drama content within the app.
\end{tabular} \\
& \\
Viewer Interaction & \begin{tabular}[t]{@{}p{10cm}@{}}
Platforms use in-feed advertisements to direct users to mini-program \\
micro-dramas. Episodes are unlocked via ad-viewing or payment. \\
Beyond mini-programs, independent creators publish serialized content \\
on personal accounts and distribute through private channels. Platforms \\
encourage likes and comments. Some creators engage with audiences \\
and adjust content based on feedback.
\end{tabular} \\
& \\
Content Creation & \begin{tabular}[t]{@{}p{10cm}@{}}
No dedicated content creation platform. Content is acquired by \\
attracting creator teams and production companies to submit videos.
\end{tabular} \\
\bottomrule
\end{tabular}
\end{table*}

\begin{table*}[t]
\centering
\caption{Video Streaming Platforms}
\label{tab:video_streaming_platforms}
\begin{tabular}{@{}ll@{}}
\toprule
\textbf{Category} & \textbf{Description} \\
\midrule
Platform Name & iQiyi / Tencent Video \\
& \\
Platform Overview & \begin{tabular}[t]{@{}p{10cm}@{}}
Traditional long-form video platforms that added micro-drama \\
sections following the rise of short dramas. Micro-dramas are \\
now accessible on these conventional streaming platforms.
\end{tabular} \\
& \\
Viewer Interaction & \begin{tabular}[t]{@{}p{10cm}@{}}
Content is accessed through paid membership subscriptions, with \\
some micro-dramas available for free. Users can watch the first \\
five episodes of premium micro-dramas, after which they must \\
purchase or subscribe to continue. Users can provide feedback \\
and suggestions through likes and comments.
\end{tabular} \\
& \\
Content Creation & \begin{tabular}[t]{@{}p{10cm}@{}}
Dedicated content creation platforms and teams. Some long-form \\
video platforms maintain independent micro-drama production teams.
\end{tabular} \\
\bottomrule
\end{tabular}
\end{table*}

\end{document}